\documentclass[aps,prb,twocolumn,amsmath,amssymb,superscriptaddress,floatfix,scrartcl,eqsecnum]{revtex4-1}

\usepackage{amssymb}
\usepackage{color}
\usepackage{graphicx}
\usepackage{mathrsfs}
\usepackage[unicode=true,pdfusetitle,
bookmarks=false,colorlinks=true,citecolor=blue,urlcolor=blue,linkcolor=red]{hyperref}

\usepackage{bm}
\usepackage{braket}
\makeatletter
\def\widebar{\accentset{{\cc@style\underline{\mskip10mu}}}}
\makeatother
\makeatletter
\def\wideubar{\underaccent{{\cc@style\underline{\mskip10mu}}}}
\makeatother

\makeatletter

\@addtoreset{equation}{section}
\makeatother

\def\LL{{\cal L}}

\begin{document}

%\preprint{}

\title{Effective hydrodynamic field theory and condensation picture of topological insulators}
\author{AtMa P. O. Chan}
\thanks{The first two authors contributed equally to the work.}
\affiliation{Department of Physics and Institute for Condensed Matter Theory, University of Illinois at Urbana-Champaign, 1110 West Green St, Urbana IL 61801-3080, USA}
\author{Thomas Kvorning}
\thanks{The first two authors contributed equally to the work.}
\affiliation{Department of Physics, Stockholm University, AlbaNova University Center, SE-106 91 Stockholm, Sweden}
\author{Shinsei Ryu}
\affiliation{Department of Physics and Institute for Condensed Matter Theory, University of Illinois at Urbana-Champaign, 1110 West Green St, Urbana IL 61801-3080, USA}
\author{Eduardo Fradkin}
\affiliation{Department of Physics and Institute for Condensed Matter Theory, University of Illinois at Urbana-Champaign, 1110 West Green St, Urbana IL 61801-3080, USA}
%


%\thanks{}


\date{\today}

\begin{abstract}
While many features of topological band insulators are commonly discussed at the level of single-particle electron wave functions,
such as the gapless Dirac spectrum at their boundary, it remains elusive to develop a {\it hydrodynamic} or {\it collective} description
of fermionic topological band insulators in 3+1 dimensions.
As the Chern-Simons theory for the 2+1-dimensional quantum Hall effect,
such a hydrodynamic effective field theory provides a universal description
of topological band insulators, even in the presence of interactions,
and that of putative fractional topological insulators.
In this paper, we undertake this task by using the functional bosonization.
The effective field theory in the functional bosonization is written in terms of a two-form gauge field, which couples to
a $U(1)$ gauge field that arises by gauging the continuous symmetry of the target system (the $U(1)$ particle number conservation).
Integrating over the $U(1)$ gauge field by using the electromagnetic duality, the resulting theory describes topological band insulators
as a condensation phase of the $U(1)$ gauge theory (or as a monopole condensation phase of the dual gauge field).
The hydrodynamic description, and the implication of its duality, of the surface of topological insulators are also discussed.
We also touch upon the hydrodynamic theory of fractional topological insulators by using the parton construction.
\end{abstract}



\maketitle

\tableofcontents

\section{Introduction}

The recent discoveries of time-reversal symmetric topological band insulators in two and
three dimensions have greatly extended our understanding on topological phenomena in condensed matter physics.
\cite{HasanKane2010, QiZhang2011, HasanMoore2011, Chiu2015}
They clearly demonstrate how topological phases beyond the physics of the quantum Hall effect (quantum Hall effect)
emerges at the level of single particle physics.
It remains, however, to be understood the effects of interactions;
we need to understand topological states of matter
where electron-electron interactions are not necessarily week,
or even those topological phases that arise precisely because of strong correlations.

%As an attempt to describe topological insulators with interactions,
%possible collective or hydrodynamic descriptions of 3+1-dimensional [3+1d] topological band insulators have widely been discussed in the literature.
One possible approach to address these questions
is to develop {\it collective} or {\it hydrodynamic} descriptions of topological (band) insulators.\cite{Cho2011,Chan2013}
These coarse-grained descriptions are to be contrasted with
more microscopic descriptions
which heavily rely on free electrons or nearly free quasiparticles.
%While a lot can be learned on the physics of 3+1d topological band insulators from
%microscopic descriptions in terms of free electrons,
%just as fluid dynamics is an efficient description of a collection of a macroscopic number of
%particles, the dynamics of topological insulators may well be described in terms of collective degrees of
%freedom, rather than relying on microscopic fermionic degrees of freedom, such as electrons
%(quasi-particles).
In fact, a hydrodynamic picture was developed for the quantum Hall effect,
and involves the Chern-Simons gauge theory.
\cite{Frohlich-1991,Wen-1992,Wen-1995}
%and we have a good understanding of a quantum Hall system as a droplet of electron liquid.
%The hydrodynamic topological field theory description of the quantum Hall droplet is given by the Chern-Simons gauge theory.
Once such effective description of the low-energy physics is established, it is likely to be robust against
interactions, and has a wider range of applicability than the non-interacting microscopic system. The
Chern-Simons field theory in the context of the fractional quantum Hall effect has been used as a vital tool
to describe and predict quasiparticle statistics, ground state degeneracy, the properties of the gapless edge
states, etc. (For a review see, e.g. Ref.\ [\onlinecite{Fradkin2013}].)

The purpose of the paper is to develop a hydrodynamic effective field theory description of
topological insulators. The method of our choice is the  {\it functional bosonization} procedure.\cite{Fradkin1994,Burgess1994}
The functional bosonization is a recipe to derive an effective action,
which reproduces the correlation functions of the conserved currents (``hydrodynamic modes'') of the system.
The functional bosonization approach relies
on the gauge invariance of the original, microscopic system,
e.g., the $U(1)$ gauge invariance of the conserve electromagnetic charge.
The resulting effective field theory contains a dynamical gauge field whose gauge group is determined
by the symmetry of the microscopic system.
In this sense,
this procedure may  also be thought of as a procedure which is akin to {\it gauging},
a useful technique to study symmetry-protected topological phases in general.
\cite{LevinGu12, Ryu2012}
%The functional bosonization approach relies
%on the gauge invariance of the original, microscopic system.
%For example,
%for a system that conserves a $U(1)$ charge,
%the partition function in the presence of external $U(1)$ gauge fields $A^{\mathrm{ex}}$,
%$
% Z[A^{\mathrm{ex}}],
%$
%is invariant under
%the electromagnetic gauge transformations,
%$
%Z[A^{\mathrm{ex}}+a]
%=
%Z[A^{\mathrm{ex}}]
%$
%where
%$
%a = d \phi
%$
%is pure gauge.
%The functional bosonization is a recipe to derive an effective action,
%which reproduces the correlation functions of the current associated
%to the gauge invariance.

By making use of the  functional bosonization procedure, in
Ref.\ [\onlinecite{Chan2013}]
 a new quantum field theory
description of (both non-interacting and interacting) topological insulators in two and three dimensions was proposed. It
consists of  the BF topological field theory supplemented with an axion term. BF topological field theory\cite{Horowitz1989} has played a key role in
the description of topological phases of matter ranging from superconductors\cite{Hansson2004,Hansson2012,Hansson2015} to topological insulators.\cite{Cho2011,Chan2013}
This is a first step toward understanding and describing the fractional topological insulator in three dimensions.
As expected, the effective field theory reproduces all universal properties of topological band
insulators, such as the topological electromagnetic effect. In addition, once written in terms of
hydrodynamic degrees of freedom, there is a natural way (at least at the level of field theories) to
incorporate the effects of interactions, in particular, the fractionalization of electrons. With the working
hypothesis of  electron fractionalization (i.e. the parton construction\cite{Blok1990a}), the effective field theory
predicts, for example, the fractionalized version of the topological magnetoelectric effect, and non-trivial
ground state degeneracy when the system is put on a manifold with non-trivial topology. Such predictions
can be compared with future numerical studies and experiments. Furthermore, the field theory description
is a natural generalization of the Chern-Simons hydrodynamic field theory for the fractional quantum Hall
effect, and hints a clue to generalize important theoretical ideas, such as the particle-vortex duality,
statistical transmutation by flux attachment, parton construction, and anomalies, among others.
\cite{Swingle2011, Maciejko2010, Maciejko2012, Mcgreevy2012, Lu2012,Callan1985,RyuMooreLudwig2012}

Guided by these previous works, we further continue to develop a  hydrodynamic
description of both interacting and non-interacting   topological insulators in 3+1 space-time dimensions.
While for the case of non-interacting topological insulators,
this may be a mere rewriting of the non-interacting theory,
it would give us a theoretical framework to discuss
weakly or moderately interacting topological insulators,
and putative fractional topological insulators.
We follow the spirit of our previous work,
and try to develop understandings in terms of the hydrodynamic degrees of freedom
-- we will make use of the hydrodynamic effective field theory.
In particular, we discuss significant issues that were left out in our previous papers.

The outline of the paper and the main results are summarized as follows:

Firstly,  as noted in Ref.\ [\onlinecite{Chan2013}] (see also Ref.\ [\onlinecite{Rosenow2013}]),
the BF-theory with the axion term is not yet
written solely in terms of hydrodynamic degrees of freedom.
The theory includes a $U(1)$ gauge field $a_{\mu}$ which is not directly tied to
hydrodynamic variables (densities)
and can be thought of
as a higher dimensional analogues of ``statistical gauge fields'',
which appear in the composite particle theories of quantum Hall liquid.
(See Sec.\ \ref{Functional bosonization}).
In this paper,
we complete our mission of deriving
effective field theories written solely in terms of
hydrodynamic degrees of freedom by integrating over
the statistical gauge field
(Sec.\ \ref{Integration over the statistical gauge field}).

Along the course of implementing
these technical steps,
we will also note that the integration of the statistical gauge field
can be viewed as a procedure which effective implements
the electromagnetic duality of the Maxwell gauge field
(Sec.\ \ref{Electromagnetic duality}).
This allows us to
develop some physical picture of topological insulators;
By making a comparison with Julia-Toulouse approach to defect condensation
\cite{Julia1979},
we will show a topological band as well as trivial insulator phase can be viewed as
a Higgs phase of the statistical gauge field.
This is in analogy with the interpretation of the quantum Hall effect in the composite boson theory
where the quantum Hall liquid is viewed as arising by the
condensation of composite bosons.
(For a similar condensation picture for bosonic topological insulators,
see Refs.\ [\onlinecite{Ye2014, Ye2015}].)
%For a similar picture for bosonic topological insulators,
%where bosonic topological insulators arise due to dyon condensation,
%see Ref.\ [\onlinecite{Ye2014, Ye2015}].

Furthermore,
we will implement another aspect which was not fully discussed in the previous work, in particular
the compact nature of the gauge field, in
Section \ref{Functional bosonization with monopole gauge invariance}.
In the spirit of the functional bosonization approach,
the method of our choice to derive the hydrodynamic field theory,
we rely on the gauge invariance of the original, microscopic theories,
e.g., the $U(1)$ gauge invariance associated to the charge conservation.
The functional bosonization of Refs.\ [\onlinecite{Fradkin1994,Burgess1994}] is a recipe that allow to derive an effective action,
which reproduces the correlation functions of the current associated
to the gauge invariance.
In the presence of monopoles,
i.e., if one were interested in the response of the system to the introduction of monopoles,
the $U(1)$ gauge field must be treated as a compact variable.
The compact nature of the $U(1)$ gauge field can be made explicit by
considering the {\it monopole gauge transformations}.\cite{Affleck1989,Fradkin1991}
They are discrete two-form gauge transformations,
which originate from the arbitrariness of the location of the Dirac strings
emanating from monopoles. (See, e.g. Ref.\ [\onlinecite{Savit1980}].)
The system must be invariant under the monopoles gauge transformations
in order for the precise locations of the Dirac strings not to affect physics.
Following the spirit of the functional bosonization,
one can derive a hydrodynamic theory for the collective variables
associated to the monopole gauge invariance.
We will show how this procedure can be implemented.
Once the compact nature of the gauge field is fully incorporated,
the resulting bosonized theory
has much resemblance with
the Cardy-Rabinovici theory
\cite{Cardy:1981qy}
and
the description of the condensed phase of the Abelian-Higgs model
in
Ref.\ [\onlinecite{Grigorio2011}].

Subsequently,
we will also discuss
the boundary (surface) of topological insulators in terms of the hydrodynamic
effective field theory in
Section\ \ref{Surface theory}.
As in the bulk, the statistical gauge field can be integrated over to
obtain a hydrodynamic effective field theory.
This process, as in the bulk, can be viewed as an implementation of
the electromagnetic duality,
\cite{Witten1995}
and relates two different 2+1 dimensional theories
with and without the statistical gauge field theory.
This surface duality is essentially the bosonized version of
the recently proposed duality between
the free Dirac fermion and QED$_3$ in Refs.\ [\onlinecite{Metlitski2015, WangSenthil2015,Metlitski2015b}].
In addition, the resulting hydrodynamic theory is compared
with the Fradkin-Kivelson theory in Ref.\ [\onlinecite{Fradkin1996}],
which enjoys $PSL(2,\mathbb{Z})$ duality symmetry.

Finally, in Section \ref{Fractional topological insulators},
we discuss putative fractional topological insulators by using the parton construction.
Assuming the electron fractionalization into partons,
we used the functional bosonization to derive the bulk and surface hydrodynamic theories of
fractional topological insulators.
We show the resulting theory in the bulk is
the $\mathbb{Z}_{\mathrm{K}}$ Cardy-Rabinovici theory with $\mathrm{K}>1$.
We conclude in Section\ \ref{Discussion} by discussing open problems.

\section{Functional bosonization, electromagnetic duality,
and the Julia-Toulouse approach}
\label{Functional bosonization in the bulk}

In this section, we review the functional bosonization
of $D=3+1$-dimensional topological insulators presented in Ref.\ [\onlinecite{Chan2013}].
For technical simplicity,
we will focus on topological insulators in symmetry class AIII in $D=3+1$,
characterized by an integer-valued topological invariant, the three-dimensional winding number $\nu$,
and  protected by  chiral symmetry.
This topological insulator is somewhat analogous to
the time-reversal symmetric topological insulator
in symmetry class AII, in that it supports a Dirac fermion
surface state, and has a nontrivial
axion-electrodynamics response to the external electromagnetic field.
The difference is, however, that the latter is characterized by
a $\mathbb{Z}_2$ topological invariant, rather than an integer
topological invariant.
To capture the $\mathbb{Z}_2$ nature of topological insulators in AII class,
one needs to consider a dimensional reduction from a one higher dimension\cite{QiHughesZhang2008} $D=4+1$,
whereas
topological insulators in class AIII in $D=3+1$ can be studied on its own.
An  example of topological insulators in AIII class
can be found in Ref.\ \onlinecite{Hosur2010} which discusses a lattice tight-binding model description.
Topological insulators in symmetry class DIII in $D=3+1$ can also be studied in a similar way.

\subsection{Functional bosonization}
\label{Functional bosonization}

We start from the partition function in the presence of
an external gauge field,
$
 Z[A^{\mathrm{ex}}],
$
where $A^{\mathrm{ex}}$ is an external $U(1)$ gauge field
associated
to the electromagnetic $U(1)$ gauge invariance.
%and $U^{\mathrm{ex}}_{\mu\nu}$ is an external gauge field
%associated to the monopole gauge invariance.
The partition function is invariant under
the electromagnetic gauge transformations,
\begin{align}
Z[A^{\mathrm{ex}}+a]
=
Z[A^{\mathrm{ex}}],
\quad
 \mbox{where}
 \quad
 a = d \phi.
 \label{u(1)gaugeinvariance}
\end{align}
By making use of the gauge invariance,
$
Z[A^{\mathrm{ex}}+a]
=Z[A^{\mathrm{ex}}]
$
with $a= d \phi$,
one can average the partition function over $a$:
\begin{align}
 Z[A^{\mathrm{ex}}]
&=
\mathcal{N}
 \int\mathcal{D}[a,b]
 Z[A^{\mathrm{ex}} + a]
 \nonumber \\%%%%%
&\qquad
\times \exp \Big(\frac{i}{2\pi} \int_{\mathcal{M}_4} b\wedge da\Big),
%%
% &\quad \times
% \exp{
%  \frac{i}{8\pi} \int d^4x\,
%  b_{\mu\nu}\epsilon^{\mu\nu\lambda\rho}
%  f_{\lambda \rho}[a]
% },
% e^{ i \int p\wedge (a-d\phi)}
\label{bzstart}
\end{align}
where $\mathcal{N}$ is a normalization constant,
and $\mathcal{M}_4$ is the spacetime manifold of our interest.
Here, the totally antisymmetric rank two tensor $b=(1/2)b_{\mu\nu}dx^{\mu}\wedge dx^{\nu}$
is introduced such that the integration over $b$
enforces the pure gauge condition $da=0$.
With a shift $a\to a - A^{\mathrm{ex}}$,
the partition function is given by
\begin{align}
 &Z[A^{\mathrm{ex}}]
 =
 \mathcal{N} \int \mathcal{D}\left[ a,b\right]
 Z[a]
 \nonumber \\
&\qquad
\times \exp \frac{i}{2\pi} \int_{\mathcal{M}_4} b\wedge (da-dA^{\mathrm{ex}}).
%\nonumber \\%%%%%
%&\qquad
%\times
%\exp{\frac{i}{2\pi}
%  \int d^4x\,
%  b_{\mu\nu} \epsilon^{\mu\nu\lambda\rho}
%  (
%  f[a] -f[A^{\mathrm{ex}}]
%  )_{\lambda\rho}
% }.
%e^{ i \int b\wedge d(a-A^{\mathrm{ex}})}
\label{step1noncompact}
\end{align}

So far, we have not assumed any microscopic details
except
for the electromagnetic $U(1)$ gauge invariance.
%and monopole gauge invariance.
i.e., the underlying system can be topologically trivial or non-trivial, and may or may not
 include interactions.
We now specialize to the case of non-interacting topological band insulators,
which can be described, at low energies, by a theory of free massive Dirac fermions.
In this case, the partition function $Z[a]$ can be evaluated by integrating over fermions
in the presence of background gauge fields $a$.
The effective action can be expanded in terms of the inverse band gap, and written as
\begin{equation}
Z[a] \propto \exp(- W[a]), 
\end{equation}
 where $W[a]$ has the form
\begin{align}
 W[a]
% &=
%\frac{1}{8\pi}
%\int
%d^4x\,
% \\
%&\quad
%\times
%\left[
%  \frac{4\pi}{g^2} f^{\mu\nu}[a]f_{\mu\nu}[a]
%%  \right.
%%\nonumber \\%%%%%
%%&\quad
%%\left.
%+
%\frac{i \theta}{4\pi}
%\epsilon^{\mu\nu\lambda\rho}
%f_{\mu\nu}[a] f_{\lambda\rho}[a]
%\right]
%+\cdots.
%\nonumber \\
&=
\frac{1}{g^2} \int da \wedge \star da
+
\frac{i\theta}{8\pi^2} \int da \wedge da
+
\cdots
%\nonumber \\
%&=
%\frac{\tau_2}{4\pi} \int da \wedge \star da
%+
%\frac{i \tau_1}{4\pi} \int da \wedge da
%+\cdots.
\end{align}
Here, $\star$ represents the Hodge dual, and $g$ is the effective coupling constant for the Maxwell term
and $\theta$ is the electromagnetic polarizabilty (the theta angle).
The theta angle is quantized,
$\theta = \pi \times (\mbox{integer})$
in the presence of time-reversal symmetry (AII)
or CT symmetry (AIII).

To summarize,
the bosonized partition function (in the
Euclidean signature) is given by
\begin{align}
&Z[A^{\mathrm{ex}}] =
\mathcal{N}\int \mathcal{D}[a,b]\ e^{-S[a,b]},
\end{align}
%where the Lagrangian density is given by
%\begin{align}
% \label{full bz lagrangian}
%\mathcal{L} &=
%-\frac{i}{8\pi} b_{\mu\nu} \epsilon^{\mu\nu\lambda\rho}
%(f_{\lambda\rho} [a] - f_{\lambda\rho}[A^{\mathrm{ex}}])
%\\%%%%%
%&\quad
%+
%%\frac{\alpha}{4}
%\frac{\tau_2}{8\pi}
%f^{\mu\nu}[a] f_{\mu\nu}[a]
%+
%\frac{i \tau_1}{16 \pi}
%f_{\mu\nu}[a]
%\epsilon^{\mu\nu\lambda\rho}
%f_{\lambda\rho}[a]
%+ \cdots
%\nonumber
%\end{align}
%Or,
where the effective Euclidean action $S[a,b]$ is
\begin{align}
 S[a,b] &= - \frac{i}{2\pi} \int b\wedge ( da-dA^{\mathrm{ex}})
 \nonumber \\
 &\quad
 +
 \frac{\tau_2}{4\pi} \int da \wedge \star da
 +
 \frac{i \tau_1}{4\pi} \int da \wedge da,
 \label{bz-action-with-a-and-b}
\end{align}
where
we have introduced a complex coupling  by
\begin{align}
 \tau = i \tau_2 + \tau_1 =
 i \frac{4\pi}{g^2} +  \frac{\theta}{2\pi}.
\end{align}

%In Minkowski signature, the bosonized Lagrangian is written as
%\begin{align}
%\mathcal{L} &=
%-\frac{1}{2} b_{\mu\nu} \epsilon^{\mu\nu\lambda\rho}
%(f_{\lambda \rho} [a] -f_{\lambda\rho}[A^{\mathrm{ex}}])
%\nonumber \\%%%%%
%& \quad
%- \frac{\alpha}{4}
%f^{\mu\nu}[a]f_{\mu\nu}[a]
%+
%\frac{\beta}{4}
%\epsilon^{\mu\nu\lambda\rho}
%f_{\mu\nu}[a]
%f_{\lambda\rho}[a]
%+ \cdots.
%\end{align}

These steps of the functional bosonization,
starting from the $U(1)$ gauge invariance in Eq.\ \eqref{u(1)gaugeinvariance}
to the final bosonized action of Eq. \eqref{bz-action-with-a-and-b},
were carried out in the previous paper, Ref.\ \onlinecite{Chan2013}.
We now raise two issues, which we will discuss in the rest of the paper.

\subsubsection{Role of the fields $b_{\mu\nu}$ and $a_{\mu}$}

The bosonized effective theory consists of the path integral over
two kinds of gauge fields, a vector (gauge) field $a_{\mu}$ and and an anti-symmetric tensor field (Kalb-Ramond) $b_{\mu\nu}$.
The first issue pertains to the role played by these two gauge fields in the functional bosonization.

From the functional derivative of $\ln Z[A^{\mathrm{ex}}]$
with respect to the external gauge field $A^{\mathrm{ex}}$,
one establishes the identification
$ \epsilon^{\mu\nu\rho\lambda} \partial_{\nu} b_{\rho\lambda}
$
as
the electromagnetic $U(1)$ current $j^{\mu}$
(``bosonization dictionary''),
\begin{align}
j^{\mu} := \frac{\delta}{\delta A^{\mathrm{ex}}_{\mu}} \ln Z[A^{\mathrm{ex}}]
=
\frac{1}{2\pi}
 \epsilon^{\mu\nu\rho\lambda} \partial_{\nu} b_{\rho\lambda}
\end{align}
(in the Minkowski signature).
On the other hand, the gauge field
$a_{\mu}$ does not appear to be related to any physical quantity.
In fact,
for the case of the $D=2+1$ dimensional quantum Hall effect or Chern insulators,
a comparison between  the functional bosonization
and
the composite boson theory
shows that
$a_{\mu}$
plays a role similar to the ``statistical gauge field'' of the theories of the fractional quantum Hall effect.\cite{Zhang1989,Lopez1991}
In the composite boson theory,
this statistical gauge field is introduced to
change the fermionic statistics of electrons
into bosonic statistics
to form ``composite bosons'' out of electrons.

In the composite boson theory of the quantum Hall effect\cite{Zhang1989} (both integer and fractional quantum Hall effect),
the composite bosons undergo condensation,
and, as a consequence, the statistical gauge field
acquires a mass by a Higgs mechanism.
To be more precise, the Meissner effect occurs
the combination of the statistical and the electromagnetic $U(1)$ gauge fields.
In the condensed phase,
%(the Higgs phase of the statistical gauge field $a_{\mu}$),
one can make use of the boson-vortex duality in $D=2+1$ dimensions to
rewrite the theory -- written in terms of the composite boson field and  the statistical gauge field --
into the theory written in terms of the ``vortex gauge field'' interacting with the statistical gauge field.
One can then integrate over the statistical gauge field to end up with
the single-component hydrodynamic Chern-Simons theory of the vortex gauge field.
This vortex gauge field corresponds to the gauge field $b_{\mu}$ in the functional bosonization.
%\footnote{
In passing, we note that %as a quick remark,
the composite particle approach is possible only when the magnetic field is non-zero,
and cannot be applied to the trivial band insulator,
while the functional bosonization does apply to both trivial and non-trivial band insulators as well.%}

Following our discussion in $D=2+1$,
we can interpret $a_{\mu}$ in $D=3+1$
%and $u_{\mu\nu}$
as a counterpart of the statistical gauge field $a_{\mu}$ in $D=2+1$ dimensions.
%As we have seen, $a_{\mu}$ implements a monopole attachment, which is akin to the flux attachment in $D=2+1$.
On the other hand, $b_{\mu\nu}$ is
directly related to the electric charge current, and hence it is a natural hydrodynamic variable, analogous to the hydrodynamic gauge field of the Fquantum Hall in 2+1 dimensions.\cite{Wen-1995}
Following $D=2+1$,
it is desirable to integrate over statistical gauge fields
$a_{\mu}$ to obtain the theory written solely in terms of the hydrodynamic variable.

\subsubsection{Condensation picture}

The above comparison with the composite particle theory (the composite boson theory) of the quantum Hall effect
leads to our second issue.
In the quantum Hall effect or in Chern insulators, the insulator phases are interpreted as a phase where composite bosons condense.
On the other hand,
it is not clear (yet) what is the physical picture suggested by the effective bosonized action of Eq.
\eqref{bz-action-with-a-and-b}.
It is highly desirable to establish a physical interpretation of (topological) insulator phases within the functional bosonization.
In the following, we will address these issues.

\subsubsection{Compact v.s. non-compact $U(1)$ gauge invariance}

Before proceeding,
it is important to emphasized that we have treated
both $A^{\mathrm{ex}}$ and $a$ as non-compact $U(1)$ gauge fields.
This may be justified, in the spirit of the functional bosonization, if we are interested only
in the response of the system to smooth configurations of $A^{\mathrm{ex}}$;
The bosonized action of Eq. \eqref{bz-action-with-a-and-b} is capable of describing the system's response
to smooth configurations of $A^{\mathrm{ex}}$, which does not include monopoles.
It is however well-known that
one of the defining properties of topological insulators is
their response to monopoles.
\cite{Qi2009Sci}
For this reason, it is desirable to develop the functional bosonization scheme which fully takes into account
the presence of monopoles in $A^{\mathrm{ex}}$ (and $a$ as well).
Note that, as a consequence of the presence of monopoles,
electric charges in the system are quantized in the unit of the elementary magnetic charge. This is the Dirac quantization condition.

Instead of considering monopoles in $A^{\mathrm{ex}}$,
one can impose the quantization of electric charges from the outset.
This in turn makes the gauge field $A^{\mathrm{ex}}$ (and $a$ as well) an angular variable,
and
the corresponding gauge group is compact $U(1)$.
(Note also that if the system of interest is defined on a lattice,
a compact $U(1)$ gauge field can be introduced naturally,
by defining  the gauge field on links.
However, the existence of an underlying lattice is neither necessary nor sufficient
to discuss the compact $U(1)$ gauge theory.)
Because of the compact nature of the gauge field,
discontinuous configurations of $A^{\mathrm{ex}}$ (and $a$) are allowed,
and hence monopoles exist in the compact $U(1)$ gauge theory.
%an underlying $U(1)$ gauge invariance is compact

These two mechanisms of charge quantization,
one because of the Dirac quantization condition in the presence of monopoles,
and the other in which it is enforced from the outset,
may seem logically independent.
These two points of view, however, are essentially the same.
Nevertheless, details of the bosonization procedure differ slightly
depending on which points of view we take;
if we take the gauge group to be a compact $U(1)$,
or if we merely allow the presence of monopoles in the system.
The latter point of view can be implemented as the {\it monopole gauge invariance},
as we will discuss.

%(It is not entirely necessary to have an underlying lattice to discuss monopoles, although we do need a short distance regulator to define the ``core'' of monopoles.)

The reason why we have emphasized compact v.s. non-compact nature of the gauge fields
is that this is closely related to the condensation picture of topological (as well as ordinary) insulators;
condensed phases of the $U(1)$ gauge theory, which can possibly coupled to matter fields of various kinds,
may be described as condensation of electric charges (Higgs phases),
condensation of magnetic charges (confined phases),
or condensation of both magnetic and electric charges (oblique confinement).
To be able to access and discuss these phases, one of which may describe (topological) insulator phases,
it is well-advised to keep the compact nature of the gauge fields and monopoles in $A^{\mathrm{ex}}$ and $a$.
In fact, as our analysis below will reveal,
topological as well as ordinary band insulator phases
can be identified as the Higgs phase of the gauge field $a$ or the monopole condensation phase of the dual gauge
field of $a$.

\subsubsection{The Julia-Toulouse mechanism}

In the next section,
we will integrate over the ``statistical'' gauge field $a_{\mu}$.
We first attempt this in a direct way (see below), and
then make a connection to the electromagnetic duality (S-duality).\cite{Witten1995}
%Along the course of the integration over $a_{\mu}$,
%we will find the following two key concepts quite useful:
%the electromagnetic duality (S-dualty)
While we have emphasized the importance of including compact nature of the $U(1)$ gauge field,
we will first proceed with the ``non-compact version'' of the bosonized action of Eq.\ \eqref{bz-action-with-a-and-b}:
We postpone to discuss the compact $U(1)$ gauge field
in Sec.\ \ref{Functional bosonization with monopole gauge invariance}.

As we will demonstrate, the $b$ field, once the compact nature of the gauge field is implemented, is treated as a discrete variables.
Even so, however, {\it if} defects (monopoles) of the dual gauge field condense,
the $b$ field can be treated as a continuum variable:
From the point of view of the dual gauge field, $db$ represents monopole currents
(recall that $db$ represents electric currents in terms of the original gauge field $A^{\mathrm{ex}}$ and $a$).
Once monopoles in the dual gauge field proliferate, $db$ can be treated as a continuum variable.
This is nothing but the Julia-Toulouse approach to defect condensation.
\cite{Julia1979}
In Appendix
\ref{Review: The Julia-Toulouse approach},
we give a short summary of the Julia-Toulouse approach
following the work of Quevedo and Trugenberger.
\cite{Quevedo1997}

In the next section,
we integrate over the non-compact statistical gauge field $a_{\mu}$ in Eq.\ \eqref{bz-action-with-a-and-b}.
This will reveal the electromagnetic duality,
which, together with the Julia-Toulouse approach, allows us to discuss the Higgs phase as well as the confined phase of
the theory with a compact $U(1)$ gauge field $a$.
In fact, the  BF coupling in the effective action of Eq.
\eqref{bz-action-with-a-and-b},
by construction, enforces $da=0$ everywhere,
which is indicative of the Higgs phase.
The compact nature of the $U(1)$ gauge fields and monopoles will be discussed in
Sec.\ \ref{Functional bosonization with monopole gauge invariance},
where the $b$ field is treated as a discrete variable.
By making a comparison with the Cardy-Ravinovici theory,
\cite{Cardy:1981qy}
we will confirm the condensation picture
even when the compact nature of the $U(1)$ gauge field is taken into account.

\subsection{Integration over the statistical gauge field}
\label{Integration over the statistical gauge field}

Since  the Euclidean action of Eq. \eqref{bz-action-with-a-and-b}
is quadratic,
%:
%\begin{align}
% \mathcal{L} = \frac{-i}{8\pi} b_{\mu\nu} \epsilon_{\mu\nu\lambda\rho} f_{\lambda\rho}[a]
% +
% \frac{\tau_2}{8\pi} f_{\mu\nu}[a] f_{\mu\nu}[a]
% +
% \frac{i \tau_1}{16 \pi} \epsilon_{\mu\nu\lambda\rho} f_{\mu\nu} f_{\lambda\rho}
%\end{align}
the integration over the gauge field $a_{\mu}$
can be done by using its equation of motion:
\begin{align}
 \frac{i}{2} \partial_{\lambda} b_{\mu\nu}
 \varepsilon^{\mu\nu\lambda\kappa}
 +
 \tau_2 \partial_{\lambda} f_{\kappa \lambda} [a]
 +
 \frac{i \tau_1}{2}
 \varepsilon^{\mu\kappa\lambda\rho} \partial_{\mu} f_{\rho\lambda}[a]
 =0,
\end{align}
where $f[a]$ is the field strength of $a$.
While the last term is identically zero (the Bianchi identity), we will keep this term to be consistent with the derivation using the electromagnetic duality.
This equation of motion can be solved by
\begin{align}
 &
 b_{\mu\nu}
 =
 \tau_1 f_{\mu\nu}
 +
  \tau_2
  \frac{i}{2}
  \varepsilon^{\mu\nu\lambda\rho}f_{\lambda\rho}. 
%\nonumber \\
%&
%\frac{i}{2} \varepsilon^{\mu\nu\lambda\rho}b_{\lambda\rho}
%=
%-\tau_2 f_{\mu\nu} + \tau_1
%\frac{i}{2} \varepsilon^{\mu\nu\lambda\rho} f_{\lambda\rho}.
\end{align}
This  solution, however, is not unique as one could add any term which vanishes when acted with the operator
$\varepsilon^{\mu\nu\lambda\kappa} \partial_{\lambda}$.
Hence, starting from the solution given above, one can generate the family of solutions
\begin{equation}
b_{\mu\nu} \to b_{\mu\nu} + \partial_{\mu}v_{\nu} - \partial_{\nu}v_{\mu},
\end{equation}
since
\begin{equation}
  \varepsilon^{\mu\nu\lambda\kappa} \partial_{\lambda} (\partial_{\mu}v_{\nu} - \partial_{\nu}v_{\mu})
 =0.
 \end{equation}
To eliminate $a_{\mu}$, we plug the solution of the equation of motion into the action to obtain
\begin{align}
S
&=
\frac{\tilde{\tau}_2}{4\pi}
\int_{\mathcal{M}_4}
(b + dv)\wedge \star (b+dv)
\nonumber \\
&\quad
+\frac{-i \tilde{\tau}_1}{4 \pi}
\int_{\mathcal{M}_4}
(b+dv)
\wedge
(b+dv)
\label{211}
\end{align}
%\begin{align}
%\mathcal{L}
%&=
%%16 \pi^2 \Big[
%\frac{\tilde{\tau}_2}{8\pi}
%(b_{\mu\nu}+f_{\mu\nu}[v])
%(b_{\mu\nu}+f_{\mu\nu}[v])
%\nonumber \\
%&\quad
%+\frac{-i \tilde{\tau}_1}{16 \pi}
%(b_{\mu\nu}+f_{\mu\nu}[v])
% \epsilon_{\mu\nu\lambda\rho}
%(b_{\lambda\rho}+f_{\lambda\rho}[v])
%%\Big]
%\end{align}
where we introduced the dual coupling as
\begin{align}
 \tilde{\tau} = i\tilde{\tau}_2 + \tilde{\tau}_1 = -\frac{1}{\tau}
\end{align}
where
\begin{align}
\tilde{\tau}_1 = -\frac{\tau_1}{\tau^2_1+\tau^2_2},
\quad
\tilde{\tau}_2 = \frac{\tau_2}{\tau^2_1+\tau^2_2},
\end{align}
are, respectively, the real and the imaginary parts of the dual coupling.

%\textcolor{blue}{Note the factor of $16\pi^2$ and $-\tilde{\tau}_1$ instead of $\tilde{\tau_1}$.}

\subsection{Electromagnetic duality}
\label{Electromagnetic duality}

We will now give more transparent derivation of the hydrodynamic action \eqref{211} by invoking the electromagnetic duality.
In particular, we will show the one-form $v$ can be interpreted as a dual gauge field to $a_{\mu}$.
%We will now show the one-form $v$ can be interpreted as a dual gauge field to $a_{\mu}$.
%This can be seen more explicitly once we invoke the well known electromagnetic duality.

\subsubsection{Review of the electromagnetic duality in the vacuum}
As a warm up, let us follow
Ref.\ \onlinecite{Witten1995}
and review the electromagnetic duality of the Maxwell theory in the vacuum
which is described by the Euclidean action
%\begin{align}
% \label{Maxwell with monopole gauge fields}
%&\mathcal{L} =
%%-\frac{i}{2} b_{\mu\nu} \epsilon^{\mu\nu\lambda\rho}
%%(f_{\lambda \rho} [a] - u_{\lambda\rho})
% \frac{\tau_2}{8\pi}
%f^{\mu\nu}[a]
%f_{\mu\nu}[a]
%+
%\frac{i \tau_1}{16\pi}
%f_{\mu\nu}[a]
%\epsilon^{\mu\nu\lambda\rho}
%f_{\lambda\rho}[a].
%\end{align}
\begin{align}
S&=
\frac{\tau_2}{4\pi} \int da \wedge \star da
+
\frac{i \tau_1}{4\pi} \int da \wedge da.
\label{Maxwell}
\end{align}
To this end,
the Maxwell action is expanded by introducing a
two-form gauge field $u_{\mu \nu}$ and a one-form gauge field $v_\mu$ as
%\begin{align}
% \label{Maxwell with monopole gauge fields}
%&\mathcal{L} =
%%-\frac{i}{2} b_{\mu\nu} \epsilon^{\mu\nu\lambda\rho}
%%(f_{\lambda \rho} [a] - u_{\lambda\rho})
%\frac{i}{8\pi}
%f_{\mu\nu}[v] \epsilon^{\mu\nu\lambda\rho}
%u_{\lambda\rho}
%\nonumber \\%%%%%
%&\quad \quad
%+ \frac{\tau_2}{8\pi}
%(f^{\mu\nu}[a]-u^{\mu\nu}) (f_{\mu\nu}[a] -u_{\mu\nu})
%\nonumber \\%%%%%
%&\quad \quad
%+
%\frac{i \tau_1}{16\pi}
%(f_{\mu\nu}[a] -u_{\mu\nu})
%\epsilon^{\mu\nu\lambda\rho}
%(f_{\lambda\rho}[a]-u_{\lambda\rho}).
%\end{align}
\begin{align}
S&=
\frac{i}{4\pi} \int dv \wedge u
\nonumber \\
&\quad
+
\frac{\tau_2}{4\pi} \int (da-u) \wedge \star (da-u)
\nonumber \\
&\quad
+
\frac{i \tau_1}{4\pi} \int (da-u) \wedge (da-u).
\end{align}
%Here, $u_{\mu\nu}$ is a two-form gauge field.
In addition to its invariance under
the electromagnetic $U(1)$ gauge transformations, this theory is also invariant under
the following two-form gauge invariance
\begin{align}
 a_{\mu} &\to a_{\mu} + \xi_{\mu},
 \nonumber \\%%%%%
 u_{\mu\nu} & \to u_{\mu\nu} + \partial_{\mu}\xi_{\nu} -
 \partial_{\nu}\xi_{\mu}.
\end{align}

The extended theory is equivalent to the Maxwell theory,
as can be seen upon integrating over $v$ to set $du=0$.
%\footnote{
%\textcolor{red}{
%N.B.
%The integral over $v$ consists of a discrete sum over
%dual line bundles $\tilde{L}$
%and for each $\tilde{L}$
%a continuous integral over connections on
%$\tilde{L}$.
%The continuous part of the integral gives a
%delta function setting $du = 0$.
%The sum over $\tilde{L}$ then gives a further delta function setting
%the periods of $v$ to be integral multiples of $2\pi$.
%But the two-form gauge invariance
%permits one
%(in a fashion that is unique up to an ordinary gauge transformation)
%to set
%$v = 0$ precisely if $dv = 0$ and
%$v/2\pi$ has integral periods.
%So after integrating over $V$,
%we gauge $v$ to zero,
%reducing the extended gauge invariance to ordinary gauge invariance,
%and reducing the Lagrangian
%to the original Maxwell Lagrangian.
%Then we can gauge away $u$ to obtain
%the original Maxwell Lagrangian (supplemented with the axion term).}}
%\begin{align}
% \label{Maxwell} %\mathcal{L} &=
% \frac{\alpha}{4}
%f^{\mu\nu}[a]f_{\mu\nu}[a]
%+
%\frac{i\beta}{4}
%\epsilon^{\mu\nu\lambda\rho}
%f_{\mu\nu}[a]
%f_{\lambda\rho}[a].
%\end{align}
Alternatively,
we can first gauge away $a_{\mu}$
by using a two-form gauge transformation,
$a_{\mu}\to a_{\mu} + (-a_{\mu})$
and
$u_{\mu\nu}\to u_{\mu\nu} - \partial_{\mu} a_{\nu}+\partial_{\nu} a_{\mu}$,
and consider
\begin{align}
S &=
\int_{\mathcal{M}_4}
\left[
\frac{i}{2\pi} dv \wedge u
+
\frac{\tau_2}{4\pi}
u \wedge \star u
+
\frac{i \tau_1}{4\pi}
u \wedge u
\right].
\end{align}
%\begin{align}
%&\mathcal{L} =
%\frac{i}{8\pi} f_{\mu\nu}[v] \epsilon^{\mu\nu\lambda\rho}
%u_{\lambda\rho}
%+ \frac{\tau_2}{8\pi}
%u^{\mu\nu}u_{\mu\nu}
%+
%\frac{i \tau_1}{16\pi}
%\epsilon^{\mu\nu\lambda\rho}
%u_{\mu\nu} u_{\lambda\rho}.
%\end{align}
One can then integrate over the two-form gauge field $u_{\lambda\rho}$ to get
the dual action,
\begin{align}
S =
\int_{\mathcal{M}_4}
\left[
\frac{\tilde{\tau}_2}{4\pi}
dv\wedge \star dv
+
\frac{i \tilde{\tau}_1}{4\pi}
dv\wedge dv
\right].
\end{align}
%\begin{align}
% \mathcal{L}
% &=
%\frac{\tilde{\tau}_{2}}{8\pi}
%f_{\mu\nu}[v] f_{\mu\nu}[v]
% +
% \frac{i \tilde{\tau}_{1}}{16\pi}
%  f_{\mu\nu}[v]
% \epsilon_{\mu\nu\lambda\rho}
%  {f}_{\mu\nu}[v],
%\end{align}
We have thus established the duality
(electromagnetic duality or S-duality)
\begin{align}
 a_{\mu}\leftrightarrow v_{\mu},
 \quad
 \tau
 \leftrightarrow
 \tilde{\tau} = -\frac{1}{\tau}.
 \label{eq:S-duality}
\end{align}
By combining the S-duality of Eq.\ \eqref{eq:S-duality} with the periodicity of the theta angle,
\begin{align}
\tau \to \tau = \tau + n,
\quad
n \in \mathbb{Z},
\label{eq:periodicity}
\end{align}
one can generate the full $SL(2,\mathbb{Z})$ (actually, $PSL(2,\mathbb{Z})$) group
of duality transformations,
which consists of the following set of {\em modular} transformations
\begin{align}
\tau \to
\frac{a\tau + b}{c\tau+d},
\quad
a,b,c,d\in \mathbb{Z},
\quad
ad-bc=1.
\end{align}
%These transformations form the modular group.

\subsubsection{Integrating over the statistical gauge field}

We now go back to our bosonized Lagrangian,
and integrate over the statistical gauge field.
Our bosonized action
differs from the
Maxwell theory, Eq.\ \eqref{Maxwell},
by the presence of $b_{\mu\nu}$.
Following the discussion on the electromagnetic duality,
we introduce two-form and one-form gauge fields,
$u_{\mu\nu}$, $v_{\mu}$,
\begin{align}
&S=
-\frac{i}{2\pi} \int b\wedge (da-u)
+
\frac{i}{2\pi} \int dv\wedge u
\nonumber \\%%%%%
&\quad \quad
+ \frac{\tau_2}{4\pi}
\int (da-u) \wedge \star (da-u)
\nonumber \\%%%%%
&\quad \quad
+
\frac{i \tau_1}{4\pi}
\int
(da-u) \wedge (da-u).
\end{align}
%\begin{align}
%&\mathcal{L} =
%-\frac{i}{8\pi} b_{\mu\nu} \epsilon^{\mu\nu\lambda\rho}
%(f_{\lambda \rho} [a] - u_{\lambda\rho})
%+
%\frac{i}{8\pi}
%f_{\mu\nu}[v] \epsilon^{\mu\nu\lambda\rho}
%u_{\lambda\rho}
%\nonumber \\%%%%%
%&\quad \quad
%+ \frac{\tau_2}{8\pi}
%(f^{\mu\nu}[a]-u^{\mu\nu}) (f_{\mu\nu}[a] -u_{\mu\nu})
%\nonumber \\%%%%%
%&\quad \quad
%+
%\frac{i \tau_1}{16\pi}
%(f_{\mu\nu}[a] -u_{\mu\nu})
%\epsilon^{\mu\nu\lambda\rho}
%(f_{\lambda\rho}[a]-u_{\lambda\rho}).
%\end{align}
Integration over
$v$ and $b$
sets $du=0$ and also $da - u=0$.
and hence the theory is in some sense
trivial
since after substituting these, the action vanishes identically.

Even in the presence of $b$,
the duality transformation presented above
can still be carried out and one obtains
\begin{align}
 \label{final-hydro-theory}
S
 &=
 \frac{i}{2\pi} \int b \wedge dA^{\mathrm{ex}}
 -\frac{i}{2\pi}
 \int (b+dv) \wedge U^{\mathrm{ex}}
 \nonumber \\%%%%%
&\quad
+
\frac{\tilde{\tau}_2}{4\pi}
\int (b+dv)\wedge \star (b+dv)
 \nonumber \\%%%%%
 & \quad
 +
 \frac{i \tilde{\tau}_1 }{4\pi}
 \int
 (b+dv)\wedge (b+dv),
\end{align}
%\begin{align}
% \label{final hydro theory}
% \mathcal{L}
% &=
% \frac{i}{8\pi} b_{\mu\nu}\epsilon_{\mu\nu\lambda\rho}
% f_{\lambda\rho}[A^{\mathrm{ex}}]
% \nonumber \\%%%%%
% &\quad
% -\frac{i}{8\pi}
% ( f_{\mu\nu}[v]+b_{\mu\nu})
% \epsilon_{\mu\nu\lambda\rho}
% U^{\mathrm{ex}}_{\lambda\rho}
% \nonumber \\%%%%%
%&\quad
%+
%\frac{\tilde{\tau}_2}{8\pi}
% \left(f_{\mu\nu}[v]+b_{\mu\nu}\right)
% \left(f_{\mu\nu}[v]+b_{\mu\nu}\right)
% \nonumber \\%%%%%
% & \quad
% +
% \frac{i \tilde{\tau}_1 }{16\pi}
% \big( f_{\mu\nu}[v] +b_{\mu\nu}\big)
% \epsilon_{\mu\nu\lambda\rho}
% \big( {f}_{\lambda\rho}[v]+{b}_{\lambda\rho}\big),
%\end{align}
where we have reinstated the external sources.
Here $U^{\mathrm{ex}}$ is the external monopole gauge field, which will be discussed in detail
in the next section.
We have thus eliminated statistical gauge fields
$a_{\mu}$ and $u_{\mu\nu}$
and express the theory in terms hydrodynamic degrees of
freedom.
A similar theory is presented, for example,
in Refs.\ [\onlinecite{Diamantini2011,Keyserlingk2012,Kapustin2014}].

%\textcolor{red}{
%Does the final action reproduce the current-current correlation function
%correctly ?}

Our final hydrodynamic theory,
Eq.\ \eqref{final-hydro-theory},
written in terms of $v$ and $b$,
is fully gapped.
This should be expected since our original theory is trivial
in the sense that upon integration over $b$ and $v$,
it sets $du=0$ and also $da - u=0$.
After substituting these, the action simply vanishes.
%(This should be compared with the ordinary
%Maxwell theory without the $b$ field.
%Witten only had $v$, and hence integration over $v$ switches off $u$,
%but the $U(1)$ Maxwell theory is still dynamical.
To see why the theory is fully gapped, we first note that
the hydrodynamic theory
of Eq.\ \eqref{final-hydro-theory}
is invariant under 2-form gauge transformations
$b_{\mu\nu}\to b_{\mu\nu}+\partial_{\mu}\xi_{\nu}-\partial_{\nu}\xi_{\mu}$
and $v_{\mu} \to v_{\mu} - \xi_{\mu}$.
By making use of this gauge invariance,
we can gauge away the one-form gauge field $v_\mu$ to obtain an action
in terms of $b_{\mu\nu}$. The resulting theory
is clearly gapped with no propagating degrees of freedom.
If one wishes, it would also be possible to
add a gauge invariant kinetic term for $b_{\lambda\rho}$,
$
\Lambda^{-2}
%( \partial_{\mu}b_{\nu\lambda} +\partial_{\nu}b_{\lambda\mu} +\partial_{\lambda}b_{\mu\nu} )^2
db\wedge \star db
$.
One then derive a (two-form) analogue of the massive
Klein-Gordon equation with the mass
$
%\propto \Lambda \sqrt{ \mathsf{q}^2 + \mathsf{p}^2} =
\propto
\Lambda |\tilde{\tau}|$.
The action of Eq.\ \eqref{final-hydro-theory} corresponds
to the infinite mass limit $\Lambda\to \infty$.
(See Refs.\ [\onlinecite{Quevedo1997}] and [\onlinecite{Diamantini2011}].)

This is nothing but a 2-form analogue of the familiar Higgs mechanism.
A natural framework to discuss this is the so-called Julia-Toulouse approach
of defect condensations -- see below.
In superconductors,
the phase of the Cooper pair $\theta$ couples to
the (dynamical) electromagnetic $U(1)$ gauge potential $A_{\mu}$
through the kinetic term
$|\partial_{\mu} \theta + q A_{\mu}|^2$
where $q$ is the charge of the Cooper pair.
We are assuming that we are in the phase where
the Cooper pairs condense and hence their amplitude are frozen.
Taking the gauge where we remove the Cooper pair phase
$\theta\to \theta + (-\theta)$
and
$qA_{\mu} \to qA_{\mu} + \partial_{\mu}\theta$,
the kinetic term reduces to the photon mass term $A_{\mu}A^{\mu}$.

Comparing with the generic prescription of the Julia-Toulouse approach,
reviewed in Appendix \ref{Review: The Julia-Toulouse approach},
in our situation, $\phi_{h-1}=v$, $\omega = b$.
We thus ended up with the picture where
band insulator = condensation of monopoles in the dual gauge field $v_{\mu}$.
This should be equivalent to the Higgs phase of the original statistical gauge field $a_{\mu}$.
Thus, as in the $D=2+1$ quantum Hall effect, we are lead to identify the insulators
as the condensation of charges in the statistical gauge field.
Observe also that in our example the topological current is given by $J_{d-h} = \star d \omega_h= db$.
By the bosonization rule,
this may simply be identified as a charge current.
Thus, the conservation of the current,
in this interpretation, is because
we enforce the theory to be in the monopole condensation phase of $v$.

%here

\section{Functional bosonization with monopoles}
\label{Functional bosonization with monopole gauge invariance}

%While we have discussed a condensation picture of band insulators,
%%invoked a comparison with the Julia-Toulouse approach,
%it should be emphasized that we have not taken into account the
%compact nature of the $U(1)$ gauge field $A^{\mathrm{ex}}$.
%Although we have formulated the bosonization scheme in the
%continuum space-time, it can be implemented in lattice models as well, as far
%as there is an underlying $U(1)$ gauge invariance
%(i.e., the particle number conservation).
%If so, it is legitimate to think of $a_{\mu }$ (``statistical
%gauge field'') as a {\protect \it compact} $U(1)$ gauge field. I.e., magnetic
%monopoles are allowed in field configurations of $a_{\mu }$.
As advocated,
we now make a further step by developing the functional bosonization
that can deal with the compactness of $A^{\mathrm{ex}}$.
%by considering the case where $A^{\mathrm{ex}}$ has monopoles.
As discussed in Sec.\ \ref{Functional bosonization},
the compact nature of the $U(1)$ gauge field can be
incorporated by demanding the quantization of charges
or by introducing monopoles into the theory.
We will implement functional bosonization in terms both of these point of view.
The resulting bosonized theory will be compared with the Cardy-Rabinovici theory.\cite{Cardy:1981qy}

\subsection{Bosonization with compact $U(1)$}

Let us recall the starting point of the bosonization,
Eq.\ \eqref{bzstart},
in which the flat connection condition $da=0$ is imposed by the auxiliary field $b$.
Instead of imposing $da  =0$ strictly,
we can impose $da \equiv 0$ mod $2\pi$ locally (i.e., for all plaquettes if we work on a lattice).
If so, the auxiliary two-form gauge field $b_{\mu \nu}$ must be a discrete variable.
(This is standard in abelian compact gauge theories on a lattice, see, e.g.,
Refs.\ [\onlinecite{Banks-1977,Kogut-1979,Savit1980}].)
To see this, we consider the generalized Poisson summation formula:
\begin{align}
\sum_{\mathcal{N}_{D-p}} &
\exp \left(2\pi i \int_{\mathcal{M}_D}
\delta_{p} (\mathcal{N}_{D-p} )\wedge A_{D-p} \right)
\nonumber \\
&\quad
=
\sum_{\mathcal{Q}_p}
\delta ( A_{D-p} - \delta_{D-p} (\mathcal{Q}_p) ),
\label{GPI}
\end{align}
valid for an arbitrary $D-p$ form $A_{D-p}$.
Here, the delta function form $\delta_{D-p}(\mathcal{N}_p)$ is
a $D-p$ form
associated to a $d$-dimensional submanifold
of spacetime $\mathcal{M}_D$,
and
defined by the relation
\begin{align}
\int_{\mathcal{M}_D} A_p \wedge \delta_{D-p}(\mathcal{N}_p) =
\int_{\mathcal{N}_p} A_p
\end{align}
for an arbitrary $p$-form $A_p$.
Useful properties of the delta forms are summarized in Appendix.
In the generalized Poisson identity, Eq. \eqref{GPI},
the summation $\sum_{\mathcal{N}_{D-p}}$ runs over all possible $D-p$ dimensional submanifolds of $\mathcal{M}_D$.

Thus, the following sum over the discrete auxiliary field $b$
\begin{align}
\sum_{b=\delta (\mathcal{M}_2)} \exp \Big(i q_e \int da\wedge b \Big)
\end{align}
enforces $da$ to be given in terms of the delta function for some manifold $\mathcal{N}_2$ as:
\begin{align}
da = 2\pi q^{-1}_e\delta(\mathcal{N}_2).
\end{align}
To summarize,
analogously to Eq.\ \eqref{bz-action-with-a-and-b},
the bosonized partition function/action is given by
\begin{align}
&Z[A^{\mathrm{ex}}] =
\mathcal{N}\int \mathcal{D}[a]\sum_{b=\delta (\mathcal{M}_2)} \exp(-S),
\end{align}
%where the Lagrangian density is given by
%\begin{align}
% \label{full bz lagrangian}
%\mathcal{L} &=
%-\frac{i}{8\pi} b_{\mu\nu} \epsilon^{\mu\nu\lambda\rho}
%(f_{\lambda\rho} [a] - f_{\lambda\rho}[A^{\mathrm{ex}}])
%\\%%%%%
%&\quad
%+
%%\frac{\alpha}{4}
%\frac{\tau_2}{8\pi}
%f^{\mu\nu}[a] f_{\mu\nu}[a]
%+
%\frac{i \tau_1}{16 \pi}
%f_{\mu\nu}[a]
%\epsilon^{\mu\nu\lambda\rho}
%f_{\lambda\rho}[a]
%+ \cdots
%\nonumber
%\end{align}
%Or,
where the Euclidean action $S$ now is
\begin{align}
 S &= - i q_e \int b\wedge ( da-dA^{\mathrm{ex}})
 \nonumber \\
 &\quad
 +
 \frac{\tau_2}{4\pi} \int da \wedge \star da
 +
 \frac{i \tau_1}{4\pi} \int da \wedge da.
\end{align}
The integration over the statistical gauge field can be done as in the non-compact case,
and we obtain the final bosonized theory
\begin{equation}
Z = \int\mathcal{D}[v] \sum_{b=\delta(\mathcal{M}_2)} \exp(-S),
\label{final bz theory}
\end{equation}
where the final form of the Euclidean action becomes
\begin{align}
S
 &=
iq_e \int b \wedge dA^{\mathrm{ex}}
 \nonumber \\%%%%%
&\quad
+
\frac{\tilde{\tau}_2}{4\pi}
\int (2\pi q_e b+dv)\wedge \star (2\pi q_e b+dv)
 \nonumber \\%%%%%
 & \quad
 +
 \frac{i \tilde{\tau}_1 }{4\pi}
 \int
 (2\pi q_e b+dv)\wedge (2\pi q_e b+dv).
\end{align}

\subsection{Bosonization with monopole gauge invariance}

\subsubsection{Review of monopole gauge invariance}

To discuss the effect of monopoles,
it is convenient to introduce
{\it a monopole gauge field}
$\Sigma_{\mu\nu}$
for
$A_{\mu}$.
\cite{Kleinert,Quevedo1997,Grigorio2011}
In the absence of monopoles, the field strength is closed, $dF_2 =0$,
and if we assume we have a manifold with trivial second homotopy group,
we can also conclude $F_2$ is exact, i.e., $F_2=dA_1$.
If, however, we allow monopoles,
\begin{align}
 dF_2 = J_{m3}
\end{align}
where $J_{m3}$ is a three-form representing the magnetic current.
For example, in the presence of a single monopole source with magnetic charge $q_m$,
it is given by
\begin{align}
 J_{m3} =
 q_m \delta_{3}( \mathcal{L}_{1}),
\end{align}
in completely analogy to the electric current 3-form,
which is given by, in terms of a world line $\mathcal{C}_1$
of point particle with charge $q_e$,
as
\begin{align}
 J_{e 3}= q_e \delta_{3}(\mathcal{C}_1).
\end{align}
%Here, the delta form $\delta_{D-p}(\mathcal{N}_p)$ is
%a $D-p$ form defined by the relation
%\begin{align}
%\int_{\mathcal{M}_D} A_p \wedge \delta_{D-p}(\mathcal{N}_p) =
%\int_{\mathcal{N}_p} A_p
%\end{align}
%for an arbitrary $p$-form $A_p$.
%Useful properties of the delta forms are summarized in Appendix.
If we consider the open submanifold
$\mathcal{N}_{2}$ which has the world-line $\mathcal{L}_1$
as its boundary, $\partial \mathcal{N}_{2} = \mathcal{L}_{1}$,
then we can write
\begin{align}
J_{m3}= q_m d\Sigma_{2},
\quad
\Sigma_{2}= \delta_{2}(\mathcal{N}_{2}),
\end{align}
where we note the formula
\begin{align}
 \delta_{D-n+1} (\partial \mathcal{N}_n) = (-1)^{n} d \delta_{D-n} (\mathcal{N}_n),
\end{align}
and hence
$ d\Sigma_{2} = d \delta_{2}(\mathcal{N}_2) = \delta (\mathcal{L}_1)$.

In the presence of monopoles, we can almost have  a connection
that has $F$ as its curvature, up to an unobservable flux tube,
\begin{align}
F_2 = dA_1 + q_m \Sigma_2,
\end{align}
where $\Sigma_2$,  the monopole gauge field, consists of unobservable flux tubes.

Physical observables, e.g. $F_2$, are invariant under a monopole gauge transformation
generated by
\begin{align}
A_1 \to A_1 +\eta_1,
\quad
\Sigma_2 \to \Sigma_2 + \sigma_2,
\end{align}
where the one-form $\eta_1$ and the two-form $\sigma_2$ are given
in terms of a 3d manifold $\mathcal{M}_3$ as
\begin{align}
\eta_1 &= q_m \delta_{1}(\mathcal{M}_{3}),\nonumber
\\
d\eta_1 &= -q_m \sigma_2,
\nonumber \\
\sigma_2 &= \delta_{2} (\partial \mathcal{M}_{3}).
\end{align}

The monopole gauge invariance, for example, directly leads to the Dirac quantization
condition of the electric and magnetic charge.
The minimal coupling between the $U(1)$ gauge field and the electric current
\begin{align}
S_{min} = q_e \int A_1 \wedge J_{e3}
\end{align}
is transformed, by monopole gauge transformations,
into
\begin{align}
 q_e \int A_1 \wedge J_{e3}
\to
 q_e \int A_1 \wedge J_{e3}
+
q_e \int \eta_1 \wedge J_{e3}.
\end{align}
Demanding the invariance of $\exp (i S_{min})$ under the monopole gauge transformations
leads to the Dirac quantization condition of the electric and magnetic charges,
\begin{align}
q_e q_m = 2\pi \times (\mbox{integer}),
\end{align}
where we used
that for arbitrary surfaces $\mathcal{M}_p$ and $\mathcal{N}_{D-p}$,
\begin{align}
\int \delta_{D-p}(\mathcal{M}_p) \wedge \delta_p(\mathcal{N}_{D-p})
=
I(\mathcal{M}_p, \mathcal{N}_{D-p})
\end{align}
where
$I(\mathcal{M}_p, \mathcal{N}_{D-p})$
is the intersection number, which is an integer.

\subsubsection{Bosonization with monopole gauge invariance}

The presence of monopoles demands
the introduction of monopole gauge invariance.
Following the spirit of the functional bosonization,
one can bosonize the conserved current associated with the monopole
gauge invariance.
Let us start from the partition function in the presence of
external gauge fields,
\begin{align}
 Z[A^{\mathrm{ex}}, U^{\mathrm{ex}}],
\end{align}
where $A^{\mathrm{ex}}$ is an external $U(1)$ gauge field
associated
to the electromagnetic $U(1)$ gauge invariance,
and $U^{\mathrm{ex}}$ is an external gauge field
associated
to the monopole gauge invariance.
The partition function is invariant under
the following two types of gauge transformations:
(i) Electromagnetic gauge transformations,
\begin{align}
& Z[A^{\mathrm{ex}}+a, U^{\mathrm{ex}}]
=
Z[A^{\mathrm{ex}}, U^{\mathrm{ex}}] ,
 \nonumber \\%%%%%
&\quad  \mbox{where}
 \quad
 a = d\phi.
 \label{gauge inv 1}
\end{align}
(ii) Monopole gauge transformations,
\begin{align}
& Z[A^{\mathrm{ex}}+ \xi, U^{\mathrm{ex}}+u]
=
Z[A^{\mathrm{ex}}, U^{\mathrm{ex}}] ,
\nonumber \\%%%%%
&\quad
\mbox{where}
\quad
d\xi+q_m u=0.
 \label{gauge inv 2}
\end{align}
Here, $u$ is a ``compact'' variable given by
\begin{align}
u = \delta_2 (\mathcal{M}_2),
\end{align}
where $\mathcal{M}_2$ has no boundaries, $\partial \mathcal{M}_2=0$.
If the topology of the spacetime is trivial,
by Poincare's lemma,
$\mathcal{M}_2$ can be written as
$\mathcal{M}_2=\partial \mathcal{M}'_3$.

The details of the functional bosonization of $Z[A^{\mathrm{ex}}, U^{\mathrm{ex}}]$
with these gauge invariance are presented in Appendix  
\ref{Functional bosonization with monopole gauge invariance app}.
The final hydrodynamic theory is given by
\begin{align}
&Z[A^{\mathrm{ex}}, U^{\mathrm{ex}}]
=
\mathcal{N}
\int \mathcal{D}[b,v] \sum^{\partial \mathcal{N}_2=0}_{w= \delta (\mathcal{N}_2)}
\exp (-S),
\label{final action, monopole gauge inv.}
\end{align}
where
\begin{align}
S
&=
i
\int (q_e b+dv) \wedge q_m U^{\mathrm{ex}}
+
i q_e \int b\wedge dA^{\mathrm{ex}}
\nonumber \\
&\quad
+
\frac{\tilde{\tau}_2}{4\pi} \int \alpha \wedge \star \alpha
+
\frac{i \tilde{\tau}_1}{4\pi} \int \alpha \wedge \alpha
\end{align}
where
\begin{align}
\frac{\alpha}{2\pi} = q_e b +  dv + 2\pi q^{-1}_m w.
\end{align}
The form of the final action is almost identical to the calculations without monopoles that we did 
in Sec.\ \ref{Functional bosonization in the bulk}. 
The only modification is the appearance of
the discrete variable $w=\delta(\mathcal{N}_2)$.

\subsection{Comparison with the Cardy-Rabinovici theory}

To develop a physical interpretation of the final bosonized action,
let us make a comparison with the Cardy-Rabinovici theory.
\cite{Cardy:1981qy}
The Cardy-Rabinovici theory is defined on a four-dimensional hypercubic lattice.
Its partition function is given by
\begin{align}
 Z= \mathrm{Tr}_{\phi, n, s}
\prod_{r} \delta[ \Delta_{\mu} n_{\mu}(r) ]
 \exp(-S),
\end{align}
where
$\phi_{\mu}$ ($\mu=1,\ldots, 4$) is a compact $U(1)$ gauge field (an angular variable) defined on the links of the lattice,
and $n_{\mu}$ and $s_{\mu\nu}$ are integer-valued fields defined respectively on links and plaquettes, respectively.
%$\phi_{\mu}$ represents an Abelian gauge field.
The integer-valued two-form gauge field $s_{\mu\nu}$ amounts to allowing multivalued configurations of the gauge field.
The sum on $s_{\mu\nu}$ corresponds to a sum over topologically non-trivial configurations
with magnetic monopoles.\cite{Banks-1977} In fact, the monopole current is given explicitly by
\begin{equation}
m_{\mu} = \frac{1}{2} \epsilon_{\mu\nu\lambda\sigma} \Delta_{\nu} s_{\lambda\sigma}.
\end{equation}
where $\Delta_{\mu}$ is the lattice difference operator in the $\mu$-direction.
%so that the final term couples $m_{\mu}$ directly to the gauge field.
On the other hand, we interpret $n_{\mu}$ as the electric current of a charge field.
The discrete delta function $\delta[ \Delta_{\mu} n_{\mu}(r) ]$ enforces current conservation.
The Boltzmann weight is given by
\begin{align}
 S
&=
 -i\mathrm{K}\sum_L n_{\mu} \phi_{\mu}
+ \frac{1}{2g^2} \sum_P \Gamma_{\mu\nu} \Gamma_{\mu\nu}
\nonumber \\
&\quad
 - \frac{i\mathrm{K}\theta }{32\pi^2}\sum_{r,r'}
f(r-r')
\epsilon_{\mu\nu\lambda\sigma}
\Gamma_{\mu\nu}(r)
\Gamma_{\lambda\rho}(r'),
\end{align}
where
$\Gamma_{\mu\nu} = \Delta_{\mu} \phi_{\nu} - \Delta_{\nu} \phi_{\mu} - 2\pi s_{\mu\nu}$
is the field strength.
The second and third terms are the Maxwell and axion terms, respectively.
(The precise nature of the smearing function $f(r-r')$ is not important here.)
%The first term in the action is a Villain-type kinetic term for the gauge field.
The sum over $n_{\mu}$ has the effect of constraining $\phi_{\mu}$ to take its values restricted to  the abelian cyclic group $\mathbb{Z}_{\mathrm{K}}$,
$\phi_{\mu} = (2\pi/\mathrm{K}) k_{\mu}$.
%This sum itself is constrained by $\Delta_{\mu} n_{\mu}=0$, since $\phi_{\mu}$ can only couple to a conserved current.
Because the sum over $n_{\mu}$ is constrained, we can always add any total divergence to $\phi_{\mu}$.
Thus, the restriction to $\phi_{\mu}= (2\pi/\mathrm{K})k_{\mu}$ represents a partial fixing of the gauge.
%Finally, $f$ is a smearing function that interpolates between plaquette and its nearest perpendicular neighbours, which we shall generally ignore below.

The action is quadratic in $\phi_{\mu}$, so we may integrate it out to obtain the Coulomb gas
representation of interacting electric and magnetic currents:
\begin{align}
Z&= \mathrm{Tr}_{n,s}\,
\prod_{r} \delta[ \Delta_{\mu} n_{\mu}(r) ]
\exp (-S),
\end{align}
where
\begin{align}
S&=
+\frac{2\pi^2}{g^2} \sum_{r,r'} m_{\mu}(r) G(r-r') m_{\mu}(r')
\nonumber \\
&\quad
+ \frac{1}{2}\mathrm{K}^2 g^2 \sum_{r,r'} \tilde{n}_{\mu}(r) G(r-r') \tilde{n}_{\mu}(r')
\nonumber \\
&\quad
- i \mathrm{K} \sum_{r,r'} m_{\mu} (r) \Theta_{\mu\nu} (r-r') n_{\nu}(r').
\end{align}
Here, $G$ is the lattice Green function and
\begin{align}
\tilde{n}_{\mu}(r) = n_{\mu}(r) +\frac{\theta}{2\pi} m_{\mu}(r)
\end{align}
is the electric current, modified to include the induced electric charges of the magnetic monopoles
due to the Witten effect.
%This form of the action expresses the dynamics of the model purely in terms of its ``topological'' excitations.
The first two terms in the new action describe the Coulomb interactions of a gas of electric and
magnetic charges. The last term represents the statistical interaction (the Aharonov-Bohm effect) between 
an electric current $\mathrm{K}g n_{\mu}$ and the Dirac string of a magnetic monopole current $2\pi g^{-1} m_{\mu}$.
$\Theta_{\mu\nu}$, the antisymmetric matrix appearing in the last term, is essentially the
``angle'' between the two currents, in the plane perpendicular to $m_{\mu}$ and $n_{\nu}$.

The duality of the model can conveniently be described by the complex coupling
$\zeta = \theta/(2\pi) + i (2\pi)/(\mathrm{K}g^2)$;
Under the duality, $\zeta \to -1/\zeta$ and $\zeta \to \zeta+1$.

By comparing entropy and free energy, Cardy and Rabinovici showed that for certain parameter ranges,
there are phases in this model where a condensate of current loops carrying electric and magnetic charges in the ratio
$-p/q$ is formed. For example, setting $m=0$, we obtain the electric charge condensation.
Observe that in this case, the theta term drops out from the action.
On the other hand, setting $n=0$, we obtain the monopole condensation.
For generic values of $p$ and $q$, the resulting phase is an example of  oblique confinement.\cite{thooft-1981}

Now, coming back to our bosonized action in Eq.\ \eqref{final bz theory}, 
the dual gauge field $v$ plays the role of $\phi$, and 
the discrete variable $b$ the role of $s$.
There is no counter part of the electric current $n$ in the bosonized action.
After integrating over $v$, the bosonized action is written in terms of
the discrete monopole charge $b$.
By comparison with the Cardy-Rabinovici theory,
the phase represented by the bosonized action is
the monopole condensation of the dual gauge field $v$.
This, in turn, implies that in terms of the original gauge field $a$,
this phase is the Higgs phase of $a$.

On the other hand,
the bosonized action \eqref{final action, monopole gauge inv.}
can be compared with the
similar field theory
presented in Ref.\ \onlinecite{Grigorio2011}
In Ref.\ \onlinecite{Grigorio2011},
the Higgs phase of the Abelian-Higgs model,
where electric charges condense,
in the presence of monopoles,
is described by the action
\begin{align}
&Z(J_g)
 = \int \mathcal{D}[B_2] \sum'_{\tilde{\Sigma}_g = \delta (\mathcal{B}_2)}
\exp (i S),
\nonumber\\
&S=
\int_{\mathcal{M}_4}
\Big\{
-\frac{1}{2e^2 v^2}
dB_2 \wedge \star dB_2
-\frac{1}{2} B_2\wedge \star B_2
\nonumber \\
&\qquad \qquad
+ g B_2\wedge \tilde{\Sigma}_g  \Big\},
\end{align}
where $\tilde{\Sigma}_g$ corresponds to the monopole gauge field for the electromagnetic $U(1)$ gauge field.
The first term in the action is the kinetic term for the two-form gauge field $B_2$, which, for our purpose, can be simply dropped. 
We can make a correspondence $b\leftrightarrow B$,
$w\leftrightarrow \tilde{\Sigma}_g$.
In our case, $w$ is the monopole gauge field of the dual gauge field.
This in turns means we are in the presence of the dual of the electric charge condensation
-- i.e. monopole condensation.

\section{Surface theory}
\label{Surface theory}

In this section, we will develop
a hydrodynamic theory that describes the $D=2+1$ dimensional
surface of 3+1-dimensional topological insulators.
%In this section, we develop a hydrodynamic description for the $2+1$ D
%surface of a $3+1$ D topological insulator by the functional bosonization of gapless Dirac fermions. 
%We obtain the surface spectrum from the hydrodynamic theory and show that it is a gapless theory. 
%Then we show that the modular transformation of the surface theory is analogous to that in the bulk theory. 
%Finally, we discuss the current-current correlation at the surface and its transformation under the modular group.}

\subsection{Functional bosonization on the surface}

Here, we derive the hydrodynamic description for the surface of a topological insulator by the functional bosonization. The surface of a $3+1$ D non-interacting topological insulator hosts gapless Dirac fermions which are described, schematically, by the lagrangian:
\begin{align}
\LL=\sum_{a=1}^{N_f}
\bar{\psi}_a i (\partial_{\mu} - i A^{\mathrm{ex}}_{\mu} )\gamma^{\mu}\psi_a,
\label{2+1 d Dirac lagrangian}
\end{align}
where $N_f$, the number of surface Dirac fermion flavors, is determined from the bulk topological invariant. 
In the following, we will focus on the case where the chemical potential is exactly at the Dirac point. One can try to apply the functional bosonization recipe developed in the preceding sections
(see also Ref.\ [\onlinecite{Rosenow2013}])
to the surface Dirac fermion theory:
\begin{align}
Z[A^{\mathrm{ex}}] &=\mathcal{N}\int\mathcal{D}[a]
\sum_{b=\delta (\mathcal{N}_2)}
\nonumber \\
&
\quad 
\times 
Z[a]
\exp \left[i \int b\wedge (da-dA^{\mathrm{ex}})\right],
\label{surface-step-1}
\end{align}
where
two one-form $U(1)$ gauge fields $b_{\mu}$ and $a_{\mu}$ are introduced.
Here  $b$ is  a compact (discrete) variable (see the comment below).
While the formal step leading to   Eq.\ \eqref{surface-step-1}
is completely identical to the corresponding step in the bulk bosonization,
compared with the functional bosonization in the gapped bulk,
one cannot organize the integration over
the gapless surface fermions
in terms of the inverse gap expansion.
Nevertheless, one can develop a systematic expansion if the number flavors $N_f$ of surface massless fermions is large.
(See also Refs.\ \onlinecite{Moreno2013, Muteeb2015}.)

To the leading order in the $1/N_f$ expansion,
the effective action $W[a]$,
related to the fermion partition function as $Z[a] \propto \exp ( -W[a])$,
is given by
\begin{align}
W[a] &= \frac{g}{4\pi} \int d^3x\, f_{\mu\nu}[a] (\partial^2)^{-1/2}f^{\mu\nu}[a]
\nonumber \\
&
\quad
+
\frac{f}{4\pi}
\int d^3x\,
\varepsilon^{\mu\nu\lambda} a_{\mu}\partial_{\nu} a_{\lambda} +\cdots
\end{align}
where $\partial^2$ is the Laplacian in 2+1 dimensional Euclidean space-time,
and $f$ and $g\sim N_f$ are parameters.

Summarizing, within the large $N_f$ expansion, the resulting bosonized theory can be written as
\begin{align}\label{bosonsurface}
S
 &=
 i
 \int d^3x\,
 b_{\mu}
 \epsilon^{\mu\nu\lambda}
 \left(
 f_{\nu\lambda}[a]-f_{\nu\lambda}[A^{\mathrm{ex}}]
 \right)
 \nonumber \\
 &\quad
 +
 \frac{1}{2}
 \int d^3x\,
 a_{\mu} D^{\mu\nu} a_{\nu},
\end{align}
where
\begin{align}
 D^{\mu\nu}
 &=
 \frac{1}{2\pi}
 \left[
  {g} (\partial^{2})^{-1/2} P^{\mu\nu}
  +
  {f} \epsilon^{\mu\alpha\nu} \partial_{\alpha}
 \right],
 \nonumber \\%%%%%
P^{\mu\nu}
&= - \partial^2 g^{\mu\nu} + \partial^{\mu}
\partial^{\nu}.
\end{align}
Here $g^{\mu\nu}$ is the metric of $\partial\mathcal{M}_4$.

Observe that the
kinetic term of the statistical gauge field $a_{\mu}$,
generated by integrating over the gapless fermions,
is non-local in spacetime.
It is this non-locality that prevents condensation
in the language of the bosonized theory.
This is necessary for the internal consistency of the functional bosonization
since once the statistical gauge field is ``Higgsed'', the bosonized theory is gapped,
whereas the original surface theory is gapless.
In turn, since we do not expect the condensation,
it is preferred to work with the discrete hydrodynamic variable $b$.
This should of course be contrasted with the Julia-Toulouse where
the discrete nature of $b$ is immaterial once the condensation happens.

Upon integrating over $a_{\mu}$,
we obtain the following effective action for the gauge field $b_\mu$,
\begin{align}\label{hydrosurface}
S
 &=
i\int d^3x\, b_{\mu}\epsilon^{\mu\nu\lambda} f_{\nu\lambda}[A^{\mathrm{ex}}]
 +
 \frac{1}{2}\int d^3x\,  b_{\mu}\tilde{D}^{\mu\nu} b_{\nu},
\end{align}
where  $\tilde{D}^{\mu \nu}$ is the operator
\begin{equation}
 \tilde{D}^{\mu\nu}
 =
 \frac{1}{2}
 \frac{1}{2\pi}
 \left[
  \tilde{g} (\partial^{2})^{-1/2} P^{\mu\nu}
  +
  \tilde{f} \epsilon^{\mu\alpha\nu} \partial_{\alpha}
 \right],
\end{equation}
and $\tilde g$ and $\tilde f$ are the dual couplings
\begin{equation}
\tilde{g}
=
\frac{g}{f^2 + g^2},
\quad
\tilde{f}
=
-
\frac{f}{f^2 + g^2}.
\end{equation}
The transformation $D^{\mu\nu}\to\tilde{D}^{\mu\nu}$
is precisely the duality transformation discussed in
Ref.\ [\onlinecite{Fradkin1996}].
Below we will give a brief review of the results of Ref.\ [\onlinecite{Fradkin1996}]. A comparison of the
the bosonized surface theory with the Fradkin-Kivelson theory will also be given shortly.

Following the discussion of the bulk electromagnetic duality,
we define a complex coupling
\begin{align}
z = f + i g.
\end{align}
%one can discuss the duality of the surface theory in the way similar to the bulk duality.
In terms of $z$,
duality is the mapping
\begin{align}
z \to \tilde{z}
=
-\frac{1}{z}.
\end{align}
Periodicity is the mapping
\begin{align}
z\to \tilde{z} = z+n,
\quad
n \in \mathbb{Z}
\end{align}
In addition, the charge conjugation is the operation
\begin{align}
z \to \tilde{z} = -z^*.
\end{align}
These transformation can be combined to form
the modular group.
\begin{align}
z\to
\frac{az + b}{cz+d},
\quad
a,b,c,d\in \mathbb{Z},
\quad
ad-bc=1.
\end{align}
Thus, similarly to the bulk,
integration over the statistical gauge field
is correlated with the duality transformation.
The derivation of the hydrodynamic theory
on the surface thus closely parallels the derivation in the bulk.
Integration over the statistical gauge field thus entails to dualization.
This connection between the bulk and surface duality transformations
is essentially what was observed by Witten in Ref.\ [\onlinecite{Witten2003}].

It should be noted that in the above discussion the duality relates two different theories.
Starting from the theory with two gauge fields $a$ and $b$,
integration over the statistical gauge field
leads to the theory solely written in terms of $b$ with the dualized coupling.
This is similar to what was observed for the 2+1-dimensional Chern-Simons theory by Witten.\cite{Witten2003}
That a duality relates two different theories is a common phenomenon in 2+1-dimensional field theories.
This should be contrasted with the electromagnetic duality of the four-dimensional $U(1)$ gauge theory
where the duality acts on the coupling constant of the theory, but do not modify the theory itself.
More recently, a duality between the free Dirac fermion and the QED has been discussed in Refs.\ [\onlinecite{Metlitski2015, WangSenthil2015}].
The 3d mirror symmetry, relating the $\mathcal{N}=2$ supersymmetric QED in 2+1 dimensions
to the so-called $XYZ$ model (the Wess-Zumino model),
is another example.
\cite{IntriligatorSeiberg1996}

To make a contact with
the recently proposed duality between the free Dirac fermion in 2+1 dimensions and QED$_3$ in Refs.\ [\onlinecite{Metlitski2015, WangSenthil2015}], 
note that the hydrodynamic surface theory in Eq.\ (\ref{hydrosurface}) is ``designed'', by the functional bosoniation receipt, 
to describe $N_f$ massless Dirac fermions at the $2+1$ dimensional surface;
The action of Eq.\ \eqref{hydrosurface}, within the large $N_f$ expansion, 
reproduces the correct effective action for $A^{\mathrm{ex}}$, which can be obtained by integrating out $b$. 
%Take $N_f=1$, the hydrodynamic surface theory describes a single massless dirac fermions coupled to $A^{ex}$. 
Now,  going back to the theory before integrating over the statistical gauge field, 
one can interpret the action of Eq.\ (\ref{bosonsurface})
as describing a ``matter field'' in terms of the gauge field $a_{\mu}$, 
which couples to a dynamical gauge field $b$. 
This matter field, following our discussion just above on the action of Eq.\ \eqref{hydrosurface}, can be interpreted as a massless Dirac fermion,
which is different from the original surface electric Dirac fermion. 
(Although one should note that $a_{\mu}$ is a compact and continuum variable, where as $b$ is a discrete variable.) 
Thus, the equivalence of the bosonized theory in Eq.\ (\ref{bosonsurface}) and the hydrodynamic theory Eq.\ (\ref{hydrosurface}) 
is exactly the particle-vortex duality discussed in Refs.\ [\onlinecite{Metlitski2015, WangSenthil2015}]. 
In other words, $db$ represents the current associated to the massless Dirac fermion, 
while $da$ is associated to the dual massless Dirac fermion.
%\textcolor{green}{(the composite fermion)}.

\subsection{Comparison with the Fradkin-Kivelson theory}

Furthermore, the hydrodynamic theory Eq.\ (\ref{hydrosurface}) is nothing but the 
theory proposed and studied in Ref.\ [\onlinecite{Fradkin1996}]. 
We now give a brief overview of the Fradkin-Kivelson theory, and compare it with the bosonized surface theory.
The Fradkin-Kivelson model is defined by
\begin{align}
Z= \sum_{\{ \ell_{\mu} \}}
\prod_{r} \delta[ \Delta_{\mu} \ell_{\mu}(r) ] \exp (-S[\ell]),
\end{align}
where $\ell_{\mu}$ is an integer-valued variable defined on a link of
a three-dimensional lattice, and represents a conserved current, i.e. the worldlines of particles in 2+1-dimensional Euclidean space-time. 
Since the currents form closed loops this theory is a theory of charged particles at charge neutrality.
In other words, the theory preserves particle-hole symmetry -- its importance has recently been discussed in 
Refs.\ [\onlinecite{Son2015, Metlitski2015, WangSenthil2015}] in the context of the half-filled Landau level and its connection to the surface of topological insulators. 
The Boltzmann weight of a configuration of loops is given by
\begin{align}
S[\ell] & =
\frac{1}{2}\sum_{r,r'} \ell_{\mu}(r) G_{\mu\nu}(r-r') \ell_{\nu}(r')
\nonumber \\
&\quad
+\frac{i}{2} \sum_{r,R} \ell_{\mu}(r)K_{\mu\nu}(r,R) \ell_{\nu}(R)
\nonumber \\
&\quad
+i \sum_{r,r'} e(r-r') \ell_{\mu}(r) A_{\mu}(r)
\nonumber \\
&\quad
+\sum_{R,R'} h(R-R') \ell_{\mu}(R) B_{\mu}(R')
\nonumber \\
&\quad
+\frac{1}{2}\sum_{r,r'} A_{\mu}(r) \Pi^0_{\mu\nu}(r,r')A_{\nu}(r').
\label{FK model}
\end{align}
Here $r$ and $R$ represent
sites on the 2+1 dimensional cubic lattice and on its dual lattice, respectively,
and
$G_{\mu\nu}$ and $K_{\mu\nu}$ are given in momentum space as
\begin{align}
G_{\mu\nu}(k) &= \frac{g}{\sqrt{k^2} } \left( \delta_{\mu\nu} - \frac{k_{\mu} k_{\nu}}{k^2} \right),
\nonumber \\
K_{\mu\nu}(k)&= i f \varepsilon_{\mu\nu\lambda} \frac{k_{\lambda}}{k^2},
\end{align}
where $f$ and $g$ are real coupling constants of the theory. 
Here $A_\mu$ and $B_\mu$ are, respectively, background gauge fields and their associated field strengths. 
The electric and magnetic charges of the particles are represented by $e$ and $h$ (in a point-split representation). 
An extension of this loop model was considered more recently by Geraedts and Motrunich,
\cite{Geraedts-2012}
in which two species of particles, each carrying either electric or magnetic charge (but not both), are introduced
(as opposed to the Fradkin-Kivelson theory, in which electric and magnetic degrees of freedom are distributed 
between the original and dual lattices, respectively, by the point splitting prescription).   

Recalling the bosonization dictionary $db\propto j$, we see that
the kernels $G_{\mu\nu}$ and $K_{\mu\nu}$ have the same structure as
 $D^{\mu\nu}$ and $\tilde{D}^{\mu\nu}$.
Thus, within the large $N_f$ expansion,
the surface of topological insulators
realizes the Fradkin-Kivelson theory in Ref.\ [\onlinecite{Fradkin1996}].
Alternatively, one can consider turning on the Coulomb interaction
by promoting $A^{\mathrm{ex}}$ into the dynamical electromagnetic $U(1)$ gauge field.
The resulting theory,
once ``projected'' to the surface,
is given by the theory discussed in Ref.\ [\onlinecite{Fradkin1996}].
(See Ref.\ [\onlinecite{Marino1993}] for a related discussion.)

From this point of view, the non-locality of the action of the loop model of Ref.\ [\onlinecite{Fradkin1996}] can be understood as a consequence of
being a theory of charged particles (with charges of both signs) that reside on a surface,  interacting through a quantized Maxwell gauge field.
The linking number represented by the odd-parity term in the action is a simply a Chern-Simons term on that surface or, equivalently,
a $\theta$ term in the 3+1-dimensional space-time.
In other words, the loop model is equivalent to Witten's modular-invariant theory\cite{Witten1995} on a four-manifold with a boundary
that represents the 2+1-dimensional space-time with a charged massless fermionic field on the boundary.
This is precisely the theory of the surface states of the 3+1-dimensional topological insulator!

This remarkable duality between 
the free Dirac fermion theory (our starting theory) 
in Eq.\ \eqref{2+1 d Dirac lagrangian},  
and 
the hydrodynamic theory whose action is given in Eq.\ \eqref{hydrosurface} (i.e., the Fradkin-Kivelson theory), 
however, must be taken with care. 
The original theory is non-interacting, whereas the Fradkin-Kivelson theory is strongly interacting. 
The situation is somewhat similar to 
the proposed duality between the free Dirac fermion in 2+1 dimensions and QED$_3$.
\cite{Metlitski2015, WangSenthil2015}
The duality may have to be regarded as a ``weak'' form of duality in the sense that there is a one-to-one correspondence 
at the level of operators (states) between in the two theories.
The functional bosonization is in fact a prescription to map the set of correlation functions in one theory into another.   
On the other hand, whether or not the ``stronger'' form of duality holds, 
in which the two theories in the IR limit are actually identical, 
is a highly non-trivial dynamical question. 
It would be possible that the free-fermion fixed point exists among the fixed points of the Fradkin-Kivelson theory.
(See the discussion below for a few interesting fixed points using the self-dual property of the Fradkin-Kivelson model.)

\subsubsection{Implications of the self duality}

We have so far discussed the duality which relates two different (2+1) dimensional theories.
However, it is possible to have a duality that acts on the same theory in $2+1$ dimensions. 
See, for example, a recent work by Xu and You. \cite{Xu2015}
%if one allows non-local terms, such as $D^{\mu\nu}$.
In Ref.\ [\onlinecite{Fradkin1996}],
the non-local 2+1-dimensional theory (the Fradkin-Kivelson theory) and its duality was shown.
(We emphasize that this duality within the Fradkin-Kivelson theory has nothing to do with the duality above discussed
for the free 2+1-dimensional Dirac fermion and the Fradkin-Kivelson theory.) 
This self-duality can be used to constrain the possible values of the
transport coefficients such as the diagonal and off-diagonal (Hall) conductance on the surface.\cite{Fradkin1996}

A duality in statistical mechanics and field theories, if exists, is a powerful
tool that allows us to make a non-perturbative prediction on the structure of phase diagrams and the
properties of the critical points even when strong interactions are present.
A famous example is the
Kramers-Wannier duality in the 2D classical Ising model that relates its high- and low-temperature phases.
\cite{Kramers-1941,Wegner-1971,Savit1980,Cardy:1981qy, Dolan2007, Shapere:1988zv}
%Duality in general, as the Kramers-Wanner duality,
%may be a powerful tool in predicting the structure of the phase diagram.
%E.g., if we assume there is a fixed point,
%the fixed point in the RG flow should be a point fixed by some group element.
%Also, once we have one fixed point,
%we automatically get a set of other fixed points as its image.

We can follow the discussion developed in Ref.\ [\onlinecite{Fradkin1996}],
where the phase diagram of
the non-local Maxwell theory
interacting with dynamical electric currents (or their dual magnetic currents)
was discussed by using the duality.
By making use of the modular symmetry,
the correlation functions and in particular
the conductivities at the modular fixed points
were exactly calculated.

Fixed points under the ${PSL}(2,\mathbb{Z})$ transformations
can be found in the following way.
We first note that the (non-abelian) modular group
${PSL}(2,\mathbb{Z})$
is generated by $S$ and $T$ with
the relation $S^2=e$ and $(ST)^3=e$.
This tells us that
${PSL}(2,\mathbb{Z})$
is essentially a free product of
$\mathbb{Z}_2$ and $\mathbb{Z}_3$.
A point fixed by $S$ can be easily found:
\begin{align}
z = i. 
\end{align}
In Ref.\ [\onlinecite{Fradkin1996}], this point is called the bosonic fixed point.
Similarly,
one can easily find a point fixed by $ST$
as
\begin{align}
 z = \frac{1}{2} + i
 \frac{\sqrt{3}}{2}
 \equiv
 \rho.
\end{align}
This is the fermionic fixed point of
 Ref.\ [\onlinecite{Fradkin1996}].
Furthermore, it is known that all the other fixed points can be found
as an image of the bosonic ($i$) or fermionic ($\rho$) fixed point.
Let $[i]$ and $[\rho]$ denote the sets of points of the upper
half plane which are the images of $i$ and $\rho$;
Also,
let $[\infty]$ denote the sets of points of the upper
half plane which are the images of $\infty$.
$[\infty]$ is the set of points of the upper half place
with $g=0$ and $f$ rational.
It can be shown that the set $\mathcal{C}$ of all points
of the upper half plane which are fixed points under
a modular transformation)
is
$ \mathcal{C}= [i]\cup [\rho]\cup [\infty]$.

By making use of the modular symmetry,
the correlation functions and in particular
the conductivities at the bosonic and fermionic fixed points
were exactly calculated in Ref.\ [\onlinecite{Fradkin1996}].

\section{Fractional topological insulators}
\label{Fractional topological insulators}

In this section, we discuss putative fractional topological insulators
using the hydrodynamic effective field theory.
We adopt the parton construction,
in which
we postulate that
electrons are fractionalized,
consist of
$\mathrm{K}$ partons,
and
each parton is in its topological insulator phase.
For each parton, we can apply functional bosonization
to derive its hydrodynamic theory.
Solving the constraints among parton densities,
we will arrive at the hydrodynamic theory of
fractional topological insulators.
See Refs.\ [\onlinecite{Maciejko2010,Maciejko2012,Swingle2011}]
for previous studies of
time-reversal symmetric
fractional topological insulators
in $D=3+1$ in terms of the parton construction.
While the parton construction may not be able to address questions regard to energetics,
it can reveal expected topological properties of fractional topological insulators.

We write down
the following action
$S= \sum_{i=1}^{\mathrm{K}} S^{(i)}
$
for partons,
%\begin{align}
%\mathcal{L}^i &=
%-\frac{1}{2} b^i_{\mu\nu} \epsilon^{\mu\nu\lambda\rho}
%(f_{\lambda \rho} [a^i] - u^i_{\lambda\rho}-
%\mathrm{K}^{-1} f_{\lambda\rho}[A^{\mathrm{ex}}]
%+\mathrm{K}^{-1} U^{\mathrm{ex}}_{\lambda\rho})
%\nonumber \\%%%%%
%&\quad
%+
%\frac{1}{2}
%f_{\mu\nu}[v^i] \epsilon^{\mu\nu\lambda\rho}
%\left(
%u^i_{\lambda\rho}
%-\mathrm{K}^{-1} U^{\mathrm{ex}}_{\lambda\rho}
%\right)
%\nonumber \\%%%%%
%& \quad
%- \frac{\alpha}{4}
%(f^{\mu\nu}[a^i]-u^{i, \mu\nu}) (f_{\mu\nu}[a^i] -u^i_{\mu\nu})
%\nonumber \\%%%%%
%& \quad
%+
%\frac{\beta}{4}
%\epsilon^{\mu\nu\lambda\rho}
%(f_{\mu\nu}[a^i] -u^i_{\mu\nu})
%(f_{\lambda\rho}[a^i]-u^i_{\lambda\rho})
%+ \cdots
%\end{align}
where
\begin{align}
S^i &=
- \int b^i\wedge (da^i - u^i- \mathrm{K}^{-1} dA^{\mathrm{ex}} +\mathrm{K}^{-1} U^{\mathrm{ex}})
\nonumber \\%%%%%
&\quad
+
\int
dv^i \wedge
\left(
u^i
-\mathrm{K}^{-1} U^{\mathrm{ex}}
\right)
\nonumber \\%%%%%
& \quad
- \frac{\tau_2}{4}
\int
(da^i-u^{i})\wedge\star (da^i -u^i)
\nonumber \\%%%%%
& \quad
+
\frac{\tau_1}{4}
\int
(da^i -u^i)
\wedge
(da^i-u^i)
+ \cdots
\end{align}
Here, the parton densities
are written in terms of
the two-form gauge fields
$
b^{i}
$
as
$j^{i}
= db^i
$,
etc.,
and are subject to the constraint
\begin{align}
&db^1 = db^2 =\cdots = db^{\mathrm{K}}= db,
\nonumber \\%%%%%
&dv^1 = dv^2 =\cdots = dv^{\mathrm{K}}= dv,
\end{align}
for all $i=1,\ldots,\mathrm{K}$.
Solving the constraint,
the resulting effective field theory is
\begin{align}
S &=
- \int b\wedge
\left[
\sum\nolimits_i
\left(
da^i - u^i
\right)
- dA^{\mathrm{ex}}
+ U^{\mathrm{ex}}
\right]
\nonumber \\%%%%%
&\quad
+
\int
dv\wedge
\left[
\sum\nolimits_i u^i
-U^{\mathrm{ex}}
\right]
\nonumber \\%%%%%
& \quad
+\sum\nolimits_{i}
\int
\Big[
- \frac{\tau_2}{4}
(da^i-u^{i})\wedge \star (da^i -u^i)
\nonumber \\%%%%%
&
\quad
\quad
+
\frac{\tau_1}{4}
(da^i -u^i)
\wedge
(da^i-u^i)
\Big].
\end{align}
We can eliminate statistical gauge fields
$a^i$ and $u^{i}$
one by one.
Gauging away $a$,
we obtain
\begin{align}
S
&=
+\int b\wedge dA^{\mathrm{ex}}
-
\int
(b+dv) \wedge U^{\mathrm{ex}}
\nonumber \\
&\quad
+ \int
\left(
b
+
dv
\right)
\wedge
\sum\nolimits_i
  u^i
\nonumber \\%%%%%
& \quad
+\sum\nolimits_i
\int
\left[
- \frac{\tau_2}{4}
u^{i}\wedge \star u^i
+
\frac{\tau_1}{4}
u^i
\wedge
u^i
\right].
\end{align}
Integrating over $u^i$, the resulting theory is
\begin{align}
S
 &=
 i \int b\wedge dA^{\mathrm{ex}}
 -i \int
 ( b+dv) \wedge U^{\mathrm{ex}}
 \nonumber \\%%%%%
&\quad
+
\frac{\tilde{\tau}_2\mathrm{K}}{2}
\int
 \left(b+dv\right)
 \wedge \star
 \left(b+dv\right)
 \nonumber \\%%%%%
 & \quad
 +
 \frac{i \tilde{\tau}_1 \mathrm{K}}{4}
 \int
 ( b+dv) \wedge (b+dv).
\end{align}
This final hydrodynamic action with fractionalization can then be related to
the Cardy-Rabinovici theory with $\mathrm{K}>1$.

More generally, we can consider
different ways to split an electron into partons.
This leads to
an analogue of the $K$-matrix theory of the fractional quantum Hall fluid:\cite{Wen-1992,Wen-1995}
\begin{align}
 \label{final hydro theory}
 S
 &=
 i \int Q_I b^I \wedge dA^{\mathrm{ex}}
 -i
 \int
 M_I ( dv^I+b^I)\wedge U^{\mathrm{ex}}
 \nonumber \\%%%%%
&\quad
+
\frac{\mathrm{K}_{IJ}}{2}
\int
 \left(b+dv\right)^I
 \wedge \star
 \left(b+dv\right)^J
 \nonumber \\%%%%%
 & \quad
 +
 \frac{i \mathrm{K}'_{IJ}}{4}
 \int
 \big( b+dv\big)^I \wedge
 \big( b+dv\big)^J.
\end{align}
where $Q^I$ is a ``charge vector''.

\section{Discussion}
\label{Discussion}

In the Landau-Ginzburg theory,
phases with spontaneous symmetry breaking -- and phase transitions between them --
are described by effective field theories built out of continuum bosonic fields associated to proper order parameters.
In discussing symmetry-protected topological phases, where we make a distinction among phases all respecting the same symmetry,
different phases cannot be described by order parameters.
Instead, {\it gauging} the symmetry, and the resulting gauge theory serves as an efficient diagnostic scheme to distinguish
different symmetry-protected topological phases.
\cite{LevinGu12, Ryu2012}
This idea is nicely materialized in the functional bosonization, in which
an effective field theory is built out of hydrodynamic variables, one for each symmetry,
which couple to corresponding gauge fields.
Following this philosophy, we have derived the hydrodynamic effective field theory of topological band insulators.
The resulting theory is found to be closely related to  the monopole-condensed phase of the Cardy-Rabinovici theory
which, in terms of the ``statistical'' $U(1)$ gauge field, corresponds to a phase with charge condensation.
We also discussed the similar hydrodynamic theory describing the surface of topological insulators,
which well compares with the Fradkin-Kivelson theory.

We close by mentioning a few issues which we have not fully explored in the paper.
One of the most interesting issues is the relevance to experiments.
The hydrodynamic mode (i.e., the $U(1)$ current) in
topological insulators can be detected by various experimental probes, e.g.,
various transport probes, and
momentum-resolved inelastic electron scattering.
\cite{Kogar2015}
Such experiments can be most interesting, in ideal systems where the bulk is truly insulating,
and when they detect the surface physics.
For example,  Kogar and coworkers\cite{Kogar2015} studied a surface plasmon mode  by
momentum-resolved electron energy-loss spectroscopy (MR-EELS).
Our bosonized surface theory, albeit within the use of the large $N_f$ expansion (qualitatively equivalent to the random phase approximation), may offer
a convenient effective field theory descriptions of the surface physics in the presence of
moderate interactions.
In particular, the duality of the surface theory pins down a particular value
of the transport coefficient, even when the surface is interacting.
It would be interesting to compare the universal results expected from the duality with future experiments.

Yet another issue is the precise connection between the bosonized surface theory and the corresponding microscopic, fermionic surface theory,
i.e., the surface Dirac fermions.
In the case of the quantum Hall effect, an important prediction from the hydrodynamic Chern-Simons theory is the existence of the gapless chiral edge state.
The vertex operators (bosonic exponentials) in the chiral edge theory then describe solitonic quasiparticle excitations, including electrons.
Such ``vertex operators'' may be constructed within the bosonized theory for the surface of 3+1-dimensional topological insulators in Sec.\ \ref{Surface theory}.
It would be then interesting to construct physical electron operators (with the Dirac dispersion) within the bosonized surface theory.
It was argued in Ref.\ [\onlinecite{Cho2011}],
based on a (different) topological field theory,
that an effective field theory can give rise to a gapless fermionic surface state with a Fermi surface.
In fact, within the non-local Maxwell theory in 2+1 dimensions, i.e., our surface theory,
we can follow the approach developed  by Marino\cite{Marino1991} to construct an fermion (electron) operator.
It should however be noted that here we focused on the situation where the chemical potential is exactly at the Dirac point (since we focused
on topological insulators in class AIII),
whereas this is not typically the case (as in Dirac surface states realized in topological insulators in symmetry class AII).

Finally, more exotic physics that may occur in the presence of strong interactions 
can also be explored within the hydrodynamic effective field theory,
including symmetry-respecting surface topological order 
\cite{Vishwanath2013, Metlitski2013a, Metlitski2013b, Burnell2013, Keyserlingk2012,Fidkowski2013, Chen2013a, Chen2013b,Wang2013a, Wang2013b, Wang2014, Bonderson2013},
and fractional topological insulators. 
For the latter, the hydrodynamic field theory in the presence of fractionalization discussed in Sec.\ \ref{Fractional topological insulators} can provide a
convenient platform to discuss, e.g., the fractionalized surface theory and its duality. 
We leave these issues for the future study.

\begin{acknowledgments}

We thank Max Metlitski and Peng Ye for useful discussion.
S.R. wishes to thank
the ESI program
``Topological phases of quantum matter''
(August 2014)
at the Erwin Schr{\"o}dinger
International Institute for Mathematical Physics in Vienna,
and the
PCTS workshop,
``Symmetry in Topological Phases''
(March 17-18, 2014)
at the Princeton Center for Theoretical Science,
where the parts of this work were presented.
The work has been supported in part
by an
INSPIRE Grant at UIUC and Stockholm University, by
the NSF under Grants
DMR-1455296 and DMR-1408713, and by the Alfred P. Sloan foundation.
\end{acknowledgments}

\appendix

\section{$\delta$-function forms}

In this Appendix, we collect useful formulas involving the $\delta$-function forms.
(The following discussion does not depend on the Euclidean/Minkowski signature of the metric.)
For an $n$-dimensional submanifold $\mathcal{N}$ of $\mathcal{M}$,
we define a $(D-n)$-form $\delta_{D-n}(\mathcal{N})$ by
\begin{align}
\int_{\mathcal{N}} A_n = \int_{\mathcal{M}} \delta_{D-n}(\mathcal{N})\wedge A_n,
\quad
\forall A_n,
\end{align}
where $A_n$ is an arbitrary $n$-form on $\mathcal{M}$.

If we flip the orientation of $\mathcal{N}$,
\begin{align}
 \delta_{D-n}(-\mathcal{N}) = -\delta_{D-n}(\mathcal{N}).
\end{align}
More generally, for oriented submanifolds $\mathcal{N}_i$,
\begin{align}
\delta\big( \sum_i c_i \mathcal{N}_i\big)
=
\sum_i c_i \delta(\mathcal{N}_i)
\end{align}
where $c_i$ is a coefficient.

Let $\mathcal{N}_1$ and $\mathcal{N}_2$ be a submanifold of $\mathcal{M}$ with
dimensions $n_1$ and $n_2$, respectively.
Define $d$
as
\begin{align}
d = n_1 + n_2 -D.
\end{align}
When $d\ge 0$, $\mathcal{N}_1$ and $\mathcal{N}_2$ can have a $d$-dimensional intersection within $\mathcal{M}$.
By properly defining an orientation, we define
$I= \mathcal{N}_1 \# \mathcal{N}_2$.
The orientation of $I$ is defined to be consistent with
\begin{align}
\delta_{D-d}(I) = \delta_{D-n_1} (\mathcal{N}_1)\wedge \delta_{D-n_1}(\mathcal{N}_2).
\end{align}

The exterior derivative of the delta form is given by
\begin{align}
\delta_{D-n+1}(\partial \mathcal{N}) = (-1)^{D-n+1} d \delta_{D-n}(\mathcal{N}).
\end{align}

\section{Details of Functional bosonization with monopole gauge invariance}
\label{Functional bosonization with monopole gauge invariance app}

In this Appendix, we derive the hydrodynamic theory
of Eq.\ \eqref{final action, monopole gauge inv.}
by the functional bosonization. 
Our starting point is the two kinds of 
gauge invariance
presented in Eqs.\ \eqref{gauge inv 1} and \eqref{gauge inv 2}. 
By making use of the monopole gauge invariance,
one can write
\begin{align}
Z[A^{\mathrm{ex}}, U^{\mathrm{ex}}]
&=
\mathcal{N}
\int \mathcal{D}[a]
\sum_{u = \delta(\partial \mathcal{M}_3)}
\delta (da+q_m u)
\nonumber \\
&\quad
\times
Z[ A^{\mathrm{ex}}+a, U^{\mathrm{ex}}+u]
\end{align}
where
$\sum_{u = \delta(\partial \mathcal{M}_3)}$
represents the sum over arbitrary submanifolds $\mathcal{M}_3$ of spacetime
with the two form $u$ given by $u=\delta(\partial \mathcal{M}_3)$.
The functional delta function
can be converted into an integral over an auxiliary field $b$,
\begin{align}
Z[A^{\mathrm{ex}}, U^{\mathrm{ex}}]
&=
\mathcal{N} \int \mathcal{D}[a,b]
\sum_{u = \delta(\partial \mathcal{M}_3)}
\nonumber \\
&\quad
\times
Z[ A^{\mathrm{ex}}+a, U^{\mathrm{ex}}+u]
\nonumber \\
&\quad
\times
\exp i q_e \int b\wedge (da+q_m u).
\end{align}
Shifting $a$ and $u$ as
$a\to a-A^{\mathrm{ex}}$ and $u\to u-U^{\mathrm{ex}}$,
%\begin{align}
%Z[A^{\mathrm{ex}}, U^{\mathrm{ex}}]
%&=
%\mathcal{N}\int \mathcal{D}[a,b]
%\sum_{u = \delta(\partial \mathcal{M}_3)}
%\nonumber \\
%&\quad
%\times
%Z[ a, U^{\mathrm{ex}}+u]
%\nonumber \\
%&\quad
%\times
%\exp \frac{i}{2\pi}\int b\wedge (da-dA^{\mathrm{ex}}+q_m u).
%%\label{step 1}
%\end{align}
\begin{align}
&Z[A^{\mathrm{ex}}, U^{\mathrm{ex}}]
=
\mathcal{N}
\int \mathcal{D}[a,b] \sum_{u= \delta (\mathcal{M}_2)}^{\partial \mathcal{M}_2 = \mathcal{L}_1}
Z[a, u]
\nonumber \\%%%%%
&\quad \times
\exp
i q_e \int
b\wedge (da +q_m u- dA^{\mathrm{ex}}- q_m U^{\mathrm{ex}}).
\label{step1}
\end{align}
The summation over $\mathcal{M}_2$ is subjected to the constraint $\partial \mathcal{M}_2=\mathcal{L}_1$,
where $\mathcal{L}_1$ is related to the external monopole gauge field $U^{\mathrm{ex}}$ as
$dU^{\mathrm{ex}} = \delta(\mathcal{L}_1)$.

Equation \eqref{step1}
is the analog of Eq.\ \eqref{step1noncompact} in the presence of monopoles,
and
is the starting point of the functional bosonization.
Before proceeding,
we note,
instead of imposing $d\xi + q_m u =0$ strictly,
we can impose $d\xi + q_m u \equiv 0$ mod $2\pi/q_e$
for all plaquette if we work on a lattice.
If so, the auxiliary field $b$ must a discrete variable.
By making use of the generalized Poisson identity,
the sum
\begin{align}
\sum_{b=\delta (\mathcal{M}_2)} \exp i q_e \int (da+q_m u)\wedge b
\end{align}
enforces $da+q_m u$ is given in terms of the delta function for some manifold $\mathcal{N}_2$:
\begin{align}
da+q_m u = 2\pi q^{-1}_e\delta(\mathcal{N}_2).
\end{align}
Here, we recall the Dirac quantization condition
%$
%q^{-1}_m = 2\pi q_e n.
%$
\begin{equation}
q_m q_e=2\pi n
\end{equation}
where $n \in \mathbb{Z}$.
The continuum v.s. discrete summation over $b$
depends on whether we assume the presence of an underlying lattice or not,
but, ultimately, this is immaterial. %does not matter so much.
In the following, we consider the continuum summation over $b$,
but it is always possible replace it with its discrete counter part.

Specializing  now to topological insulators,
the fermion partition function $Z[a,u]$ is given by
\begin{equation}
Z[a,u] \propto \exp (- W[a,u]),
\end{equation}
 where $W[a,u]$ now is
\begin{align}
 W[a,u]
&=
\frac{\tau_2}{4\pi} \int (da+q_m u) \wedge \star (da+q_m u)
\nonumber \\
&\quad
+
\frac{i \tau_1}{4\pi} \int (da+q_m u) \wedge (da+q_m u)
+\cdots.
\end{align}
Then,
the partition function is written as
\begin{align}
&Z[A^{\mathrm{ex}}, U^{\mathrm{ex}}]
=
\mathcal{N}
\int \mathcal{D}[a,b] \sum_{u= \delta (\mathcal{M}_2)}^{\partial \mathcal{M}_2 = \mathcal{L}_1}
\exp (-S[a,b,u]),
\end{align}
where
the action $S[a,b,u]$ is given by
\begin{align}
S[a,b,u]&=
-i\int q_e b\wedge (da +q_m u- dA^{\mathrm{ex}}- q_m U^{\mathrm{ex}})
\nonumber \\
&\quad
+
\frac{\tau_2}{4\pi} \int (da+q_m u) \wedge \star (da+q_m u)
\nonumber \\
&\quad
+
\frac{i \tau_1}{4\pi} \int (da+q_m u) \wedge (da+q_m u)
+\cdots.
\end{align}

The sum over $\mathcal{M}_2$ with the constraint $\partial \mathcal{M}_2=\mathcal{L}_1$
can be converted into an unrestricted sum over $\mathcal{M}_2$,
by introducing an auxiliary field $v$,
\begin{align}
&Z[A^{\mathrm{ex}}, U^{\mathrm{ex}}]
=
\mathcal{N}
\int \mathcal{D}[a,b,v] \sum_{u= \delta (\mathcal{M}_2)}
\exp (-S[a,b,u,v]),
\end{align}
where
\begin{align}
S[a,b,u,v]
&=
-i
\int dv \wedge ( q_m u -q_m U^{\mathrm{ex}})
\nonumber \\
&\quad
-i q_e \int b\wedge (da +q_m u- dA^{\mathrm{ex}}- q_m U^{\mathrm{ex}})
\nonumber \\
&\quad
+
\frac{\tau_2}{4\pi} \int (da+q_m u) \wedge \star (da+q_m u)
\nonumber \\
&\quad
+
\frac{i \tau_1}{4\pi} \int (da+q_m u) \wedge (da+q_m u)
+\cdots.
\label{final action before integrating stat. fields}
\end{align}

We now proceed to the integrate over the statistical gauge fields
$a$ and $u$, which do not couple to the external fields
$A^{\mathrm{ex}}$ and $U^{\mathrm{ex}}$.
We first integrate over $u$ and then $a$.
To this end, we introduce an auxiliary field $\alpha$,
with which the action is given by
%\begin{align}
%&\quad
%+
%\frac{\tau_2}{4\pi} \int (da+q_m u) \wedge \star (da+q_m u)
%\nonumber \\
%&\quad
%+
%\frac{i \tau_1}{4\pi} \int (da+q_m u) \wedge (da+q_m u)
%\nonumber \\
%&
%\Rightarrow
%\int \frac{i}{2\pi} (da+q_m u) \wedge \alpha
%\nonumber \\
%&\quad
%+
%\frac{\tilde{\tau}_2}{4\pi} \int \alpha \wedge \star \alpha
%+
%\frac{i \tilde{\tau}_1}{4\pi} \int \alpha \wedge \alpha
%\end{align}
%The total action with the auxiliary field $\alpha$ is
\begin{align}
S
&=
-i
\int dv \wedge ( q_m u -q_mU^{\mathrm{ex}})
\nonumber \\
&\quad
-i q_e \int b\wedge (da +q_m u- dA^{\mathrm{ex}}- q_m U^{\mathrm{ex}})
\nonumber \\
&\quad
+
\frac{i}{2\pi}
\int
(da+q_m u) \wedge \alpha
\nonumber \\
&\quad
+
\frac{\tilde{\tau}_2}{4\pi} \int \alpha \wedge \star \alpha
+
\frac{i \tilde{\tau}_1}{4\pi} \int \alpha \wedge \alpha.
\end{align}
By integrating over the auxiliary field $\alpha$, one
recovers the action Eq.\eqref{final action before integrating stat. fields}.
By making use of the generalized Poisson identity,
the sum over $u$ results in
\begin{align}
&
\sum_{u=\delta (\mathcal{M}_2)}
\exp
\int
i
\left( q_m  dv +  q_mq_e b - \frac{q_m}{2\pi} \alpha
\right)\wedge u
\end{align}
which enforces the constraint
\begin{align}
%&\quad
%q_m q_e dv +  q_mq_e b - q_m \alpha
%=
%2\pi \delta(\mathcal{N}_2)
%\nonumber \\
%&\quad
%\Rightarrow
\frac{\alpha}{2\pi} = q_e b +  dv + q^{-1}_m (2\pi) \delta(\mathcal{N}_2).
\end{align}
Then, after integrating over $u$, we find
%\begin{align}
%S
%&=
%-\frac{i}{2\pi}
%\int dv \wedge (  -U^{\mathrm{ex}})
%\nonumber \\
%&\quad
%-\frac{i}{2\pi} \int b\wedge (da - dA^{\mathrm{ex}}- q_m U^{\mathrm{ex}})
%\nonumber \\
%&\quad
%+
%\frac{i}{2\pi}
%\int
%da \wedge (b+q^{-1}_m dv + q^{-1}_m (2\pi)^2 \delta(N_2))
%\nonumber \\
%&\quad
%+
%\frac{\tilde{\tau}_2}{4\pi} \int \alpha \wedge \star \alpha
%+
%\frac{i \tilde{\tau}_1}{4\pi} \int \alpha \wedge \alpha
%\end{align}
\begin{align}
S
&=
i
\int dv \wedge q_m U^{\mathrm{ex}}
\nonumber \\
&\quad
+i q_e \int b\wedge (dA^{\mathrm{ex}}+ q_m U^{\mathrm{ex}}).
\nonumber \\
&\quad
+
\frac{i}{2\pi}
\int
da \wedge ( dv + q^{-1}_m (2\pi) \delta(\mathcal{N}_2))
\nonumber \\
&\quad
+
\frac{\tilde{\tau}_2}{4\pi} \int \alpha \wedge \star \alpha
+
\frac{i \tilde{\tau}_1}{4\pi} \int \alpha \wedge \alpha.
\end{align}
Now, integrating over $a$ sets
\begin{equation}
 q_e d^2 v + q^{-1}_m (2\pi) d \delta(\mathcal{N}_2) =0,
 \end{equation}
which implies $\mathcal{N}_2=\partial \mathcal{M}_3$.
This completes the derivation  of Eq.\ \eqref{final action, monopole gauge inv.}.

\subsection{An alternative derivation of Eq.\ \eqref{step1}}

Equation \eqref{step1} can be derived in an alternative way as follows:
Bosonizing the electromagnetic current,
with a shift $a\to a - A^{\mathrm{ex}}$,
the partition function is given by
\begin{align}
 &Z[A^{\mathrm{ex}}, U^{\mathrm{ex}}]
 =
 \mathcal{N} \int \mathcal{D}\left[ a,b\right]
 Z[a, U^{\mathrm{ex}}]
\nonumber \\%%%%%
&\qquad
\times
\exp
\frac{i}{2\pi } \int_{\mathcal{M}_4} b\wedge (da-dA^{\mathrm{ex}}).
%  \int d^4x\,
%  b_{\mu\nu} \epsilon^{\mu\nu\lambda\rho}
%  ( f[a] -f[A^{\mathrm{ex}}] )_{\lambda\rho} }.
%e^{ i \int b\wedge d(a-A^{\mathrm{ex}})}
\end{align}
We now make use of the monopole gauge invariance of
$Z[a, U^{\mathrm{ex}}]$,
$Z[a, U^{\mathrm{ex}}]
=
Z[a+\xi, U^{\mathrm{ex}}+u]$
with
$d\xi+q_m u=0$.
Following the bosonization of the electromagnetic $U(1)$ current above,
we can average over $u$ and $\xi$ as
\begin{align}
&Z[a, U^{\mathrm{ex}}]
=
\mathcal{N}
\sum_{u,\xi}
Z[a+\xi, U^{\mathrm{ex}}+u]
%e^{ i \int q\wedge  (u - d\xi) }
\end{align}
where the summation is over arbitrary boundaryless
surfaces $\mathcal{M}_2$ and $\xi$, such that
$u= \delta(\mathcal{M}_2)$ and $d\xi +q_m u=0$.
Implementing the latter condition by introducing an auxiliary field $q$,
\begin{align}
&Z[a, U^{\mathrm{ex}}]
=
\mathcal{N} \int \mathcal{D}[q]\sum_{u,\xi}
Z[a+ \xi, U^{\mathrm{ex}}+u]
%e^{ i \int q\wedge  (u - d\xi) }
\nonumber \\%%%%%
&\qquad
\times
\exp \frac{i}{2\pi}
\int_{\mathcal{M}_4} q\wedge (d\xi+q_m u),
%\exp{
% \frac{i}{2} \int d^4x\, q_{\mu\nu} \epsilon^{\mu\nu\lambda\rho}
%(u_{\lambda\rho} - \partial_{\lambda}\xi_{\rho} +\partial_{\rho}\xi_{\lambda})
%},
\end{align}
where
$\sum_{u,\xi}$ is over {\it arbitrary} manifolds $\mathcal{M}_2$ and $\mathcal{N}_3$,
respectively:
\begin{align}
\sum_{u} = \sum_{u\in \delta(\mathcal{M}_2)},
\quad
\sum_{\xi} = \sum_{\xi\in q_m \delta(\mathcal{N}_3)}.
\end{align}
The integration over $q$ enforces the constraint
$d\xi+q_m u=0$,
and reduces $\sum_{u,\xi}$ to
\begin{align}
 \sum_{u \in \delta(\partial \mathcal{N}_3)}.
\end{align}
I.e., the summation over $u_2$ is now given in terms of
{\it boundaryless} manifolds.
(In fact, $u$ doesn't have to be discrete, and can be replaced by $\int \mathcal{D}[u]$.)

With a shift
$a\to a-\xi$,
the total partition function is
\begin{align}
 & Z[A^{\mathrm{ex}}, U^{\mathrm{ex}}]
 =
 \mathcal{N}
\int \mathcal{D}[a,b,q]
\sum_{u,\xi}
Z[a, U^{\mathrm{ex}}+u]
\nonumber \\
&\qquad \times
\exp \frac{i}{2\pi}\int b\wedge (da-d\xi-dA^{\mathrm{ex}})
\nonumber \\
&\qquad \times
\exp \frac{i}{2\pi}\int q\wedge (d\xi+q_m u).
\end{align}
We first consider the summation over $\xi$:
\begin{align}
&
\sum_{\xi \in q_m \delta (\mathcal{N}_3)}
\exp
\frac{i}{2\pi} \int (- b +q)\wedge d\xi
\nonumber \\
&
\quad
=
\sum_{\xi \in q_m \delta (\mathcal{N}_3)}
\exp
\frac{i}{2\pi} \int d(b -q)\wedge \xi.
\end{align}
By using the generalized Poisson identity,
\begin{align}
&
\sum_{\mathcal{N}_{D-p}}
\exp 2\pi i \int_{\mathcal{M}_D}
\delta_{p} (\mathcal{N}_{D-p} )\wedge A_{D-p}
\nonumber \\
&\quad
=
\sum_{\mathcal{Q}_p}
\delta ( A_{D-p} - \delta (\mathcal{Q}_p) )
\end{align}
the summation over $\xi$ sets
\begin{align}
q_m d(b-q) = (2\pi)^2 \delta(\mathcal{Q}_1).
\end{align}
While at this stage $\mathcal{Q}_1$ appears to be arbitrary,
since $d^2=0$, $\partial \mathcal{Q}_1$ should be zero.
(We have used the same discussion in defining the monopole gauge field $\Sigma$.)
By the Poincar\'e lemma,
the 2-form $b-q$
can be written in terms of
a two-dimensional surface $\mathcal{P}_2$ and
a one-form $v$ as
\begin{align}
b - q = q^{-1}_m (2\pi)^2 \delta(\mathcal{P}_2) - q^{-1}_m (2\pi)^2 dv.
\end{align}
(The minus here is purely a convention.)
We can introduce $w$ as
\begin{align}
w_2= \delta_2(\mathcal{P}_2),
\quad
dw_2 = \delta(\mathcal{Q}_1).
\end{align}
By eliminating $q$,
\begin{align}
&Z[A^{\mathrm{ex}}, U^{\mathrm{ex}}]
=
\mathcal{N}
\int \mathcal{D}[a,b,v] \sum_{u,w}
Z[a, U^{\mathrm{ex}}+u]
\nonumber \\%%%%%
&\quad \times
\exp
\frac{i}{2\pi}
\int
b\wedge (da - dA^{\mathrm{ex}}+  q_m u)
 \nonumber \\%%%%%
 &\quad \times
 \exp
 -2\pi i \int
 (dv-w) \wedge  u.
\end{align}
Comments:
(i) $w$ couples only to $u$ through:
\begin{align}
\exp - 2\pi i \int w\wedge u.
\end{align}
This is always 1, so one can drop the sum over $w$.
(If $u$ to be taken as a continuous variable rather than discrete,
this summation sets $u$ to be a discrete variable.)
(ii)
$v$ couples only to $u$ through:
\begin{align}
\exp - {2\pi i}  \int dv\wedge u.
\end{align}
Integration over $v$ sets $du=0$, and hence, summation over $u$
becomes over boundaryless surfaces, as expected.
Thus, the functional integral reduces to
\begin{align}
&Z[A^{\mathrm{ex}}, U^{\mathrm{ex}}]
=
\mathcal{N}
\int \mathcal{D}[a,b] \sum_{u\in \delta (\partial \mathcal{M}_3)}
Z[a, U^{\mathrm{ex}}+u]
\nonumber \\%%%%%
&\quad \times
\exp
\frac{i}{2\pi}
\int
b\wedge (da - dA^{\mathrm{ex}} + q_m u).
\end{align}
This is nothing but Eq.\ \eqref{step1}.

\section{Review: The Julia-Toulouse approach}
\label{Review: The Julia-Toulouse approach}

In this appendix,
we give a short summary of the Julia-Toulouse approach following the work of
Quevedo and Trugenberger
\cite{Quevedo1997},
which conveniently describes the $h$-form generalization of the Higgs mechanism
($h=0,1,\ldots$).
We consider the class of field theories that
contain {\it compact} $(h-1)$-form $\phi_{h-1}$
with (generalized) gauge invariance under transformation
\begin{align}
\phi_{h-1} \to \phi_{h-1} + d\lambda_{h-2}.
\label{original gauge inv}
\end{align}
The dynamics of the field $\phi_{h-1}$ may be described by the following generic low-energy effective action:
\begin{align}
 S &=
 \int \frac{(-1)^{h-1}}{e^2}
 d\phi_{h-1} \wedge (\star d\phi_{h-1}) + \kappa\int \phi_{h-1} \wedge \star j_{h-1}
\end{align}
where $e$ and $\kappa$ are coupling constants,
$j_{h-1}$ describes a conserved (tensor) current of fields whose dynamics
is governed by the action $S_M$ (not included above).
The spacetime dimension is $D=d+1$.
For example:
when $h=2$, $\phi_{1}=:A$ is a one-form,
the theory is nothing but the compact QED:
\begin{align}
 S &=
 \int \frac{1}{e^2}
 dA \wedge (\star dA) + \kappa \int A \wedge \star j_{1}
\end{align}
When $h=1$, on the other hand, $\phi_0 =: \phi$ is a scaler field, and the action is given by
\begin{align}
 S &=
 \int \frac{1}{e^2}
 d\phi \wedge (\star d\phi) + \kappa \int \phi \wedge \star j_{0}.
\end{align}

The Julia-Toulouse approach provides a prescription to write down
an effective action describing a phase where defects in $\phi_{h-1}$ are condensed.
The effective action is given by
\begin{align}
S
&=
\int
\frac{(-1)^h}{\Lambda^2}
d\omega_{h}\wedge \star d\omega_{h}
\nonumber \\
&\quad
+
\frac{(-1)^{h-1}}{e^2}\int
(\omega_h - d\phi_{h-1}) \wedge \star (\omega_h - d\phi_{h-1})
\nonumber \\
&
\quad
+ \kappa\int  (\omega_h - d\phi_{h-1}) \wedge \star T_h,
\label{Julia-Toulouse action}
\end{align}
where
$\omega_h$ is a $h$-form gauge field,
%and $\Omega_{h+1}:=d\omega_h$.
and
$\Lambda$ is an energy scale associated to the condensation.
%\textcolor{red}{
%(Observe that this is very similar to
%Witten's manipulation.
%However, unlike Witten's
%(i) there is no auxiliary field enforcing
%$\Omega_{h+1}=0$.
%(ii) In Witten's case, the theory is conformal (Coulomb phase),
%whereas here we are interested in the phase where monopole is condensed.
%(iii) }
The guiding principle in constructing the effective Lagrangian is
the gauge invariance under the two gauge symmetries.
The first gauge invariance is the invariance under the original gauge transformation, Eq.\eqref{original gauge inv}.
In addition, we require invariance under the following gauge invariance:
\begin{align}
 \omega_h \to \omega_h + d\psi_{h-1},
 \quad
 \phi_{h-1} \to \phi_{h-1} + \psi_{h-1}.
\end{align}
Concurrently to this gauge invariance,
in the last line of the effective action of Eq. \eqref{Julia-Toulouse action},
the original conserved $(h-1)$-form current $j$ is promoted to an $h$-form current $T_h$,
which is given by $dT_h = j_{h-1}$.

The physical meanings of $\omega_h$ and the second gauge invariance are the following:
We are interested in phases where
topological defects in $\phi_{h-1}$ (which is a compact variable) condense.
These topological defects can be characterized by
an integer valued topological invariant
\begin{align}
\int_{S_h} \omega_h,
\quad
\omega_h = d\phi_{h-1},
\end{align}
where  $S_h$ is an $h$-dimensional sphere surrounding the singularity
and $\omega_h$ is the topological density.
If there is a single topological defect, $d\omega_h$ is zero almost everywhere,
but at the defect, $d\omega_h\neq 0$ (delta function peak).
For example, in the compact QED, $\omega$ is given by $\omega=dA$.
If there is a magnetic monopole, $d\omega\neq 0$ at the location of the monopole, but $d\omega$ is zero otherwise.
We further define
%\begin{align}
% \Omega_{h+1} := d\omega_h.
%\end{align}
the ``topological current'' by
\begin{align}
 J_{d-h}
 %= \star \Omega_{h+1}
 = \star (d\omega_h).
\end{align}
If there is no defect, $J_{d-h}$ is identically zero.
If there is a defect, there will be a delta function singularity.
If there are many defects, we can ``smear'' delta function singularity,
and can treat them as a constant background.
In this situation, the topological current $J_{d-h}$ is conserved.
Writing the topological current in terms of $\omega_h$ as above,
there is a redundancy:
$\omega_h\to \omega_h + d\psi_{h-1}$
gives the same topological current.
This is the origin of the emergent gauge symmetry in the condensed phase,
which we have made use of
to construct the effective action in the condensed phase.

\bibliography{reference}

\end{document}